\documentclass[
 reprint,
 amsmath,amssymb,
 aps,
]{revtex4-2}

\usepackage{graphicx}

\begin{document}

\title{$B^2$ to $B$-linear magnetoresistance due to impeded orbital motion}

\author{R. D. H. Hinlopen}
\affiliation{H. H. Wills Physics Laboratory, University of Bristol, Tyndall Avenue, Bristol BS8 1TL, United Kingdom}

\author{F. A. Hinlopen} 
\affiliation{Fudura B.V., Marsweg 5P, 8013 PD Zwolle, Netherlands}

\author{J. Ayres}
\affiliation{H. H. Wills Physics Laboratory, University of Bristol, Tyndall Avenue, Bristol BS8 1TL, United Kingdom}

\author{N. E. Hussey}
\affiliation{H. H. Wills Physics Laboratory, University of Bristol, Tyndall Avenue, Bristol BS8 1TL, United Kingdom}
\affiliation{High Field Magnet Laboratory (HFML-EMFL) and Institute for Molecules and Materials, Radboud University, Toernooiveld 7, 6525 ED Nijmegen, Netherlands}

\small
\date{\today}

\begin{abstract}
    Strange metals exhibit a variety of anomalous magnetotransport properties, the most striking of which is a resistivity that increases linearly with magnetic field $B$ over a broad temperature and field range. The ubiquity of this behavior across a spectrum of correlated metals – both single- and multi-band, with either dominant spin and/or charge fluctuations, of varying levels of disorder or inhomogeneity and in proximity to a quantum critical point or phase – obligates the search for a fundamental underlying principle that is independent of the specifics of any material. Strongly anisotropic (momentum-dependent) scattering can generate $B$-linear magnetoresistance but only at intermediate field strengths. At high enough fields, the magnetoresistance must eventually saturate. Here, we consider the ultimate limit of such anisotropy, a region or regions on the Fermi surface that impede all orbital (cyclotron) motion through them, but whose imposition can be modelled nonetheless through a modified Boltzmann theoretical treatment. Application of the proposed theorem suggests that the realization of quadratic-to-linear magnetoresistance requires the presence of a bounded sector on the Fermi surface possibly separating two distinct types of carriers. While this bounded sector may have different origins or manifestations, we expect its existence to account for the anomalous magnetotransport found in a wide range of correlated materials.
\end{abstract}

\maketitle

\section{Introduction}

It is well understood from semi-classical theories that isotropic single-band metals exhibit no magnetoresistance (MR) \cite{Pippard1989}. The introduction of a weak and $T$-independent anisotropy results in a positive MR and Kohler's scaling whereby plots of $\Delta\rho/\rho(0)$ vs.~$B/\rho(0)$ collapse onto a single curve \cite{Kohler1938}. Here, $\Delta\rho=\rho(B,T)-\rho(0,T)$ is the MR and $\rho(0,T) = \rho(0)$ is the $T$-dependent zero-field resistivity. This scaling describes the observed MR in most metals and forms one of the key successes of Boltzmann transport theory. 

Kohler's rule may be violated in systems containing multiple bands, $T$-dependent anisotropy \cite{Abdel-Jawad2006}, a separation of lifetimes \cite{Anderson1991, Chien1991lifetime} or open orbits \cite{Wu2020, Lifshitz1957}. In addition, a lack of saturation in the MR can occur in perfectly charge-neutral semimetals \cite{Ali2014}. None of these known violations, however, accounts for the quadrature MR first reported in BaFe$_2$(As$_{1-x}$P$_{x}$)$_2$ (Ba122) near its antiferromagnetic (AFM) quantum critical point (QCP) \cite{Hayes2016}. The specific form of quadratic to non-saturating and non-accidental linear MR (QLMR) found in Ba122 can be viewed either as an equivalence of $B$ and $T$, or as $B/T$ scaling:

\begin{eqnarray}
\rho &=& \rho_{0} + \sqrt{(\alpha T)^2 + (\gamma B)^2} \nonumber \\
(\rho-\rho_{0})/T &=& \alpha \sqrt{1 + (\gamma B/\alpha T)^2} \label{quadrature}
\end{eqnarray}

Here, $\alpha$ and $\gamma$ are fitting parameters used to describe the MR over the full temperature and magnetic field range studied while $\rho_{0}$ = $\rho(0,0)$ is the (extrapolated) residual resistivity. We emphasize that Eq.~\eqref{quadrature} is empirical in nature and not a theoretical form, but that where deviations have been observed, a clear universal characteristic remains \cite{Hayes2016, GiraldoGallo2018, Ayres2021}, which will be referred to as QLMR.

Since its discovery in the pnictides, QLMR has been observed in iron chalcogenides near a nematic QCP \cite{Licciardello2019Nat, Licciardello2019PRR}, in heavy fermions near a Kondo QCP \cite{Hayes2016, Weickert2006}, in CrAs near a double helical endpoint \cite{Niu2017} and in the electron- \cite{Sarkar2019} and hole-doped cuprates both inside \cite{GiraldoGallo2018} and outside \cite{Ayres2021} of the pseudogap regime. In almost all cases, $\rho(T)$ also exhibits a dominant $T$-linear dependence at low $T$ and zero field that has been linked to Planckian dissipation -- the maximum dissipation
allowed by quantum mechanics \cite{Licciardello2019Nat, Zaanen2004, Bruin2013, Legros2019, Brown2019, Cao2020}. 

The pervasiveness of QLMR among strange and quantum critical metals is striking and explanations involving quantum MR \cite{Abrikosov1998, Huynh2011}, sharp Fermi surface corners \cite{Pippard1989, Feng2019} or Zeeman splitting \cite{Hayes2016, GiraldoGallo2018, Ando2002} seem unlikely given the widely differing scattering rates, Fermi surface geometries and magnetic field orientation dependencies encountered. Realistic theoretical explanations for this behavior thus far fall into two categories. The first, based on random resistor networks \cite{Parish2003, Johnson2010}, attributes QLMR to the presence of disorder, either through real space binary distributions \cite{Boyd2019}, real space patches \cite{Patel2018}, or doping inhomogeneity \cite{Singleton2020}. The second is intrinsic and driven by cyclotron orbits in combination with nesting fluctuations or peaks in the density of states arising through hot spots \cite{Koshelev2016}, turning points \cite{Koshelev2013}, magnetic breakdown \cite{Naito1982} or van Hove singularities \cite{Grissonnanche2020}.

Beyond its striking universality, a number of fundamental challenges confront the search for a coherent explanation of QLMR. One mystery is the separation of the residual resistivity from the MR of Eq.~\eqref{quadrature}, which has proven to be more than merely suggestive. Indeed, in pnictides, chalcogenides and cuprates, large variations in $\rho_0$ appear to have limited or no influence on the magnitude of the $B$-linear slope or the characteristic (quadratic-to-linear) turnover point \cite{Licciardello2019PRR, GiraldoGallo2018, Ayres2021, Maksimovic2020}. Such insensitivity to disorder is rare in correlated electron systems \cite{Alexandradinata2020} and presents a serious challenge to theory.

Another key aspect to address is the fact that many of these systems exhibit signatures of two-component behavior in their magnetotransport, manifest as two fluids (in a multi-band material) \cite{Licciardello2019PRR}, two charge sectors (in a single-band material) \cite{Ayres2021, Culo2021} or two (or more) lifetimes \cite{Abdel-Jawad2006, Clayhold2010}. Despite being well established experimentally, it has proven difficult to determine whether these different contributions or lifetimes couple in series or in parallel, or to explain how the addition of a secondary (inelastic) scattering mechanism influences magnetotransport already at the Planckian limit. 

Besides the role of disorder and the coupling paradox, the sensitivity of the QLMR to magnetic field orientation introduces another level of complication. In the longitudinal configuration with field and current parallel to one another, QLMR is still observed in cuprates \cite{Ayres2021}, yet is absent in Ba122 \cite{Hayes2018}. The first observation is incompatible with conventional cyclotron motion in two dimensions, the second incompatible with the theory that QLMR emerges as the addition of thermal and Zeeman energy scales in a variational sense. As far as we are aware, no current theory can satisfactorily explain such symmetry differences. 

Finally, no signs of saturation of the $B$-linear MR nor of Shubnikov-de Haas oscillations have been reported (within current magnetic field ranges) in any material exhibiting quadrature scaling. Collectively, these four challenges beg the question: is there any hope of finding a \textit{universal} explanation of quadrature scaling as an alternative to Kohler's rule?

The purpose of this article is to suggest the groundwork to affirmatively answer this question by considering the essential role of pronounced anisotropy; the key postulate being that QLMR stems from a strong impedance to cyclotron motion somewhere on the Fermi surface. This impedance can result from the presence of hot spots \cite{Koshelev2016} or van Hove singularities \cite{Grissonnanche2020}, hot lines \cite{Rosch1999}, Fermi surface sectors \cite{Culo2021} or partially gapped Fermi surfaces \cite{Borisenko2008, Borisenko2009, Flicker2016} caused, e.g. by AFM or charge density wave (CDW) correlations \cite{Feng2019, Maksimovic2020}. We propose that an effective boundary of as yet unknown origin may emerge between two $k$-space separated regions within a single Fermi sheet to explain the observation of QLMR and a reduced Hall conductivity in overdoped cuprates \cite{Ayres2021, Putzke2019}. The most surprising aspect, however, is that quadrature scaling can be captured at all through orbital effects and Boltzmann theory. Despite the long standing successes of Boltzmann theory, it is by no means obvious that it could be used to describe the non-quasiparticle behavior associated with Planckian dissipation. Nevertheless, recent success in using Boltzmann theory to describe the phenomenology of strange metals \cite{Grissonnanche2020, Maksimovic2020} justifies a careful consideration. We reiterate that the focus of this paper is not towards microscopics, but rather towards establishing a phenomenology. 

In the following section, we discuss the assumptions that underpin the use of Boltzmann theory to describe impeded cyclotron motion. While Section \ref{sec:impeded} provides some insight into the consequences of Fermi surface boundaries, it lacks critical details such as the precise quadrature form and the conditions for unsaturated MR. To deal with these challenges, we provide in Section \ref{sec:specific} the simplest quantitative example of the proposed theorem using an isotropic two-dimensional (2D) Fermi cylinder with azimuthal boundaries at multiples of $\pi/2$. We study the effects of shorting and disorder in Sections \ref{sec:shorting} and \ref{sec:disorder}, respectively, while Sections \ref{sec:discussion} is reserved for discussion and outlook.

\section{Impeded cyclotron motion}
\label{sec:impeded}

In a conventional metal, the conductivity is defined as the current-current correlation over time $<\vec{j}(0)|\vec{j}(-t)>$, where the average is taken over all charge present. Given the wide
range of microscopics that could be responsible for the boundary, we will assume information about their specific origins -- while necessary for explaining why cyclotron motion is impeded -- is not relevant for describing the qualitative features of QLMR. As a result, the conductivity simplifies to the velocity-velocity correlation over time $<\vec{v}(0)\vec{v}(-t)>$ at the Fermi level and we focus here on the Lorentz contribution.

While self-evidently valid in the case of conventional quasiparticles, the applicability of Boltzmann transport theory to strange metals exhibiting non-quasiparticle behavior is not immediately apparent. Boltzmann transport theory itself is not predicated on the presence of quasiparticles, however. It has long been asserted that in the limit of long wavelengths and low frequencies, Boltzmann theory follows more generally from conservation laws \cite{Baym1961, Abrahams2003}. Recent theoretical work affirms this view more concretely through the generalization of quasiparticles to quanta \cite{Else2020_1, Else2020_2}.  The two conditions for the definition of quanta are 1) the existence of lattice translational symmetry such that a Brillouin zone is defined and 2) microscopic charge conservation, or more formally in the 2D case, the net time invariance in equilibrium of $n(\phi)$. Because the symmetry operator for translation and the charge density $n(\phi)$ for a given wave vector direction $\phi$ are connected through an effective $\vec{k}(\phi)$, this $\vec{k}(\phi)$ can be used to define an ersatz Fermi surface. We apply Boltzmann theory to the leading order out-of-equilibrium occupation level of this Fermi surface under electric fields in the relaxation time approximation. The Boltzmann transport equation incorporating anisotropic scattering can then be expressed through the Shockley-Chambers tube integral formalism (SCTIF) \cite{Shockley1950, Chambers1952}:

\begin{equation}
	\sigma_{ij} = \frac{e^2}{4\pi^3\hbar} \int_{FS} d^2k \int_0^{\infty}dt \frac{v_i(0)}{v_F(0)} v_j(-t) \exp\left(-\int_0^t \frac{dt'}{\tau_0(-t')}\right)
	\label{sigma_general_no_temperature}
\end{equation}
where FS is the Fermi surface, $\sigma$ the conductivity, $i,j\in \{x,y,z\}$ (though we will focus here on $\sigma_{xx}$ and $\sigma_{xy}$ in 2D), $e$ the elementary charge, $\hbar$ Planck's constant and $t$ time. The velocity $v$ is defined through $\nabla_k\epsilon/\hbar$ where $\epsilon$ is energy while $\tau_0$ is the anisotropic zero-field scattering rate. For simplicity, we assume an isotropic cylindrical Fermi surface. By enforcing the Fermi surface to be continuously connected \cite{Else2020_1} independent of the chemical potential, we naturally arrive at an isotropic effective mass through $\frac{1}{m^*}:=\frac{1}{\hbar^2} \frac{d^2\epsilon}{dk^2}$. 

Central to our hypothesis is the notion of impedance to orbital motion due to boundaries at discrete locations on the Fermi surface. Without magnetic field, these barriers are obsolete as no quanta encounter them. In the presence of a magnetic field, orbital motion is impeded for quanta close enough to the boundary (see Figure~\ref{fig:1fluid}a) for a schematic representation). At sufficiently high magnetic field, quanta can traverse a full sector between two boundaries. As a result, almost all charge correlation will end at the boundaries, rather than by scattering through their usual scattering channels such as disorder, phonons, magnons, other electrons, \textit{etc..}. 

Conventionally, local impedance to cyclotron motion can arise via two mechanisms: through a local reduction in $\tau_0$ (e.g. at a scattering hot spot) or through a local enhancement of $m^*$ (density of states), which we argue are equivalent in this context. In the low-field limit, the conductivity is set by the curl of the mean free path over the Fermi surface \cite{Ong1991, Harris1995}, which is given at every $k$ point by $l=v_F\tau_0\sim\tau_0/m^*$. In contrast, the high-field limit is defined by the variation of $\omega_c\tau_0$ (= $qB\tau_0/m^*$) over the Fermi surface, where $\omega_c$ is the cyclotron frequency and $q=\pm e$ the fundamental charge. In either case, the ratio $\tau_0/m^*$ is critical and boundaries arising from a local decrease in $\tau_0$ or increase in $m^*$ are deemed to be equivalent. The third possibility is impedance through a partial gapping of the Fermi surface or a suppression of spectral weight. In all cases, velocity correlation effectively terminates at specific points on the Fermi surface.

According to the relaxation time approximation, the velocity-velocity correlation decays over time as $\exp(-t/\tau_0)$ and is integrated to $t$ = $\infty$. The fundamental change introduced here is that this correlation terminates at the $k$-space boundaries, manifesting as an upper limit on the time integral in Eq.~\eqref{sigma_general_no_temperature}:

\begin{equation}
	\sigma_{ij} = \frac{e^2}{4\pi^3\hbar} \int_{FS} d^2k \int_0^{bound}dt \frac{v_i(0)}{v_F(0)} v_j(-t) \exp\left(-\frac{t}{\tau_0}\right)
	\label{sigma_general_bounded}
\end{equation}

This termination of correlation is the only non-Drude component, illustrating the minimal nature of the model. Additionally, the bound time diverges in the limit $B\rightarrow 0$ and the Drude result is recovered. In the high-field limit, $\omega_c\tau(B)$ tends towards a constant equal to the sector size $W$ in radians (see Fig.~\ref{fig:1fluid}a). In this regime, the magnetic field determines through $\omega_c\sim B$ how quickly the boundaries are reached, ultimately guaranteeing $\tau(B)\sim 1/B$. The resulting conductivity scales through the termination of the correlation function as $\sigma\propto <v(0)|v(-t)>\propto 1/B$ for both diagonal and off-diagonal elements. This scaling is distinct from $\sigma_{xx}\sim1/B^2$ and $\sigma_{xy}\sim 1/B$ in conventional theory. Through matrix inversion, we thus find simultaneous $B$-linear resistivity and Hall resistivity as limiting high-field behavior, as observed. Furthermore, the traversal time for a sector ($\sim W/\omega_c$ subject to further anisotropy) does not vary with temperature thereby ensuring that the magnitude of the $B$-linear slope is $T$-independent.

Note that this regime does not exist when the anisotropy is weak, as in the standard treatment of MR that leads to Kohler's scaling. In that circumstance, quanta can traverse the high scattering regions well before cyclotron motion between high scattering regions is possible. As a result, no effective boundary is realized and the anisotropy gradually washes out with increasing field strength, resulting in conventional high-field saturation of the MR.

Empirically, strange metals are characterised by a ubiquitous $T$-linear scattering rate often linked to the Planckian limit. Through the relation $\Delta E\Delta t\ge \hbar$ with $k_BT$ for the energy uncertainty and $\tau$ for uncertainty in time \cite{Zaanen2004}, we arrive at $\tau_0\sim 1/T$. Combining the Planckian ansatz with impeded cyclotron motion results in the observed resistivity scaling throughout the entire $(B,T)$ phase diagram. We stress, however, that impeded cyclotron motion and Planckian dissipation are two independent ingredients. In the absence of a QCP, such as in most CDW systems, impeded orbital motion can occur without Planckian dissipation. Consequently, $B/T$ scaling is seen to be only one manifestation of QLMR. Conversely, while QLMR can result from the presence of quantum critical fluctuations, it is not a prerequisite for its observation.

\section{Specific model}
\label{sec:specific}

In order to demonstrate the effect of impeded cyclotron motion on the MR, we consider here the simple example of an isotropic 2D Fermi cylinder of radius $k_F$ with tetragonal symmetry and azimuthal boundaries at multiples of $\pi/2$. No further anisotropy is introduced, the pocket can be of arbitrary size, and energy broadening through the Fermi-Dirac distribution is neglected. We emphasize that the introduction of a specific dimensionality, single-fluidity and boundary location are arbitrary and not in any way necessary for the realisation of QLMR, but a common scenario for real systems exhibiting quadrature scaling. The quanta are assumed to have charge $q=-e$, spin degeneracy and an effective mass $m^*$ which for simplicity equals their cyclotron mass $m_c$. Throughout, we will assume the field is applied along the symmetry axis and drives quanta into the boundaries through cyclotron motion with frequency $\omega_c=qB/m^*<0$. Restricting ourselves to the first quadrant, we obtain the following contribution to $\sigma_{xx}$ for azimuthal angle $\phi\in [0,\pi/2]$:

\begin{equation}
	\sigma_{xx} = \frac{e^2 k_F^2}{2\pi^2\hbar c} \int_{0}^{\pi/2} d\phi \int_0^{-\phi/\omega_c}dt 
	\frac{v_x(0)}{v_F(0)} v_x(-t)\exp(-t/\tau_0)
	\label{sxx_simple_q1}
\end{equation}
where $c$ is the $c$-axis lattice parameter. The introduction of the upper bound in the time integral here is equivalent to the introduction of a delta function in the scattering rate every $\phi = \pi/2$ in an otherwise isotropic 2D Fermi surface. The time dependence of the quantum's position on the Fermi surface is given by semi-classical equations of motion, meaning $\phi(-t)=\phi(0)+\omega_ct$. We repeat the same procedure for the second quadrant and find quadrants 1 and 3 are by symmetry equal to 2 and 4. Introducing the carrier density $n=k_F^2/2\pi c$ and repeating the procedure for hole-like carriers, we obtain:

\begin{figure}[htbp]
	\footnotesize
	\includegraphics[width=\columnwidth]{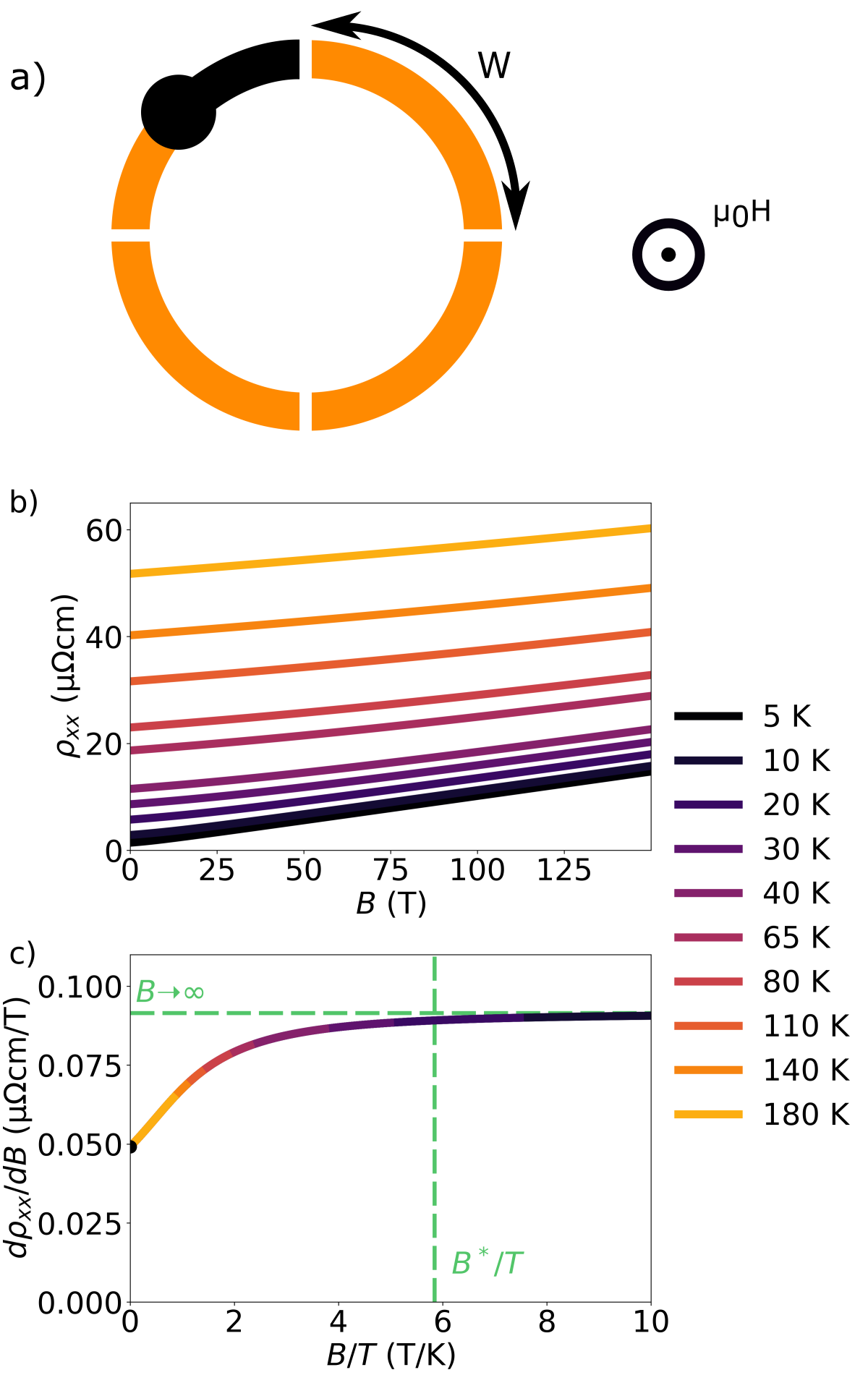}
	\caption{Quadrature scaling in a one-fluid model with a $\delta$-function impedance. a) Schematic representation of cyclotron motion on a 2D electron-like Fermi surface with four boundaries. The magnetic field direction is orthogonal to the plane and the motion represents electrons from the past or holes to the future terminating at the impedance. $W$ indicates the sector size. b) In-plane resistivity $\rho(B,T)$ computed using $n=8 \times 10^{27}$ m$^{-3}$, $c$ = 10 $\text{\AA}$, $m^*=5m_e$, parameters based loosely on a typical cuprate. The boundaries are $\delta$-functions and the $T$-dependent (isotropic) scattering rate is given by $\tau_0=\hbar/k_BT$, i.e. with no impurity scattering. c) The QLMR and scaling collapse is most evident in a derivative plot. With the infinitely sharp boundaries considered here, a residual $B$-linear term -- persisting down to the lowest fields as indicated by the black dot -- is also found. As shown elsewhere, this residual $B$-linear term is unstable to the introduction of \lq soft' boundaries or the inclusion of a second fluid. The green horizontal line represents Eq.~\eqref{1fluid_rho_highB} and the green vertical line Eq.~\eqref{Bstar}. Note that $B^*$ is considerably higher than the bending point usually assigned as the field scale through the quadrature form. Despite the simplicity of the model, the scaling collapse, the turnover scale and the magnitude of the $B$-linear slope are all found to be in good agreement with experimental observation.}
	\label{fig:1fluid} 
\end{figure}

\begin{align}
	\sigma_{xx}	= \frac{ne^2\tau_0}{m^*} \frac{1}{1+\omega_c^2\tau_0^2}
	&\left( 1 - 
	\frac{2|\omega_c|\tau_0}{\pi} \frac{1-\omega_c^2\tau_0^2}{1+\omega_c^2\tau_0^2} - 
	\right. \nonumber \\
	&\left. \quad \frac{4\omega_c^2\tau_0^2}{\pi(1+\omega_c^2\tau_0^2)} e^{-\pi/2|\omega_c|\tau_0}
	\right)
	\label{sigmaxx_analytical}
\end{align}

The same derivation can be repeated but replacing $v_x(0)v_x(-t)$ with $v_y(0)v_x(-t)$ in Eq.~\eqref{sxx_simple_q1} to obtain $\sigma_{yx}=-\sigma_{xy}$. 

\begin{align}
	\sigma_{xy} = \frac{ne^2\tau_0}{m^*} \frac{\omega_c\tau_0}{1+\omega_c^2\tau_0^2}
	&\left( 1 + 
	\frac{2}{\pi}\frac{1-\omega_c^2\tau_0^2}{1+\omega_c^2\tau_0^2}
	e^{-\pi/2|\omega_c|\tau_0} -
	\right. \nonumber \\
	&\left. \quad \frac{4}{\pi}\frac{|\omega_c|\tau_0}{1+\omega_c^2\tau_0^2}
	\right)
	\label{sigmaxy_analytical}
\end{align}

$\sigma_{xy}$ is anti-symmetric and $\sigma_{xx}$ is symmetric under both charge and time reversal symmetry (reversing the sign of $q$ or $B$), as required. The low-field behavior of $\rho_{xx}$ and $\rho_{xy}$ can be found from a Taylor expansion of the resistivity as follows:

\begin{eqnarray}
\lim_{B\rightarrow0}\rho_{xx} &=& \frac{m^*}{ne^2\tau_0} + \frac{2|B|}{ne\pi} + \mathcal{O}(B^2)
\label{rho_xx_1lfuid_low}\\
\lim_{B\rightarrow0}\rho_{xy} &=& \frac{B}{nq}\left(1+\frac{8e|B|\tau_0}{\pi m^*} \right) +\mathcal{O}(B^3)
\label{rho_xy_1lfuid_low}
\end{eqnarray} 

Here, $q=\pm e$ is the charge of a quantum. Similarly, the high-field regime is extracted using a Laurent series and results in a linear MR {\it and} Hall resistivity.

\begin{eqnarray}
	\lim_{B\rightarrow\infty}\rho_{xx} &=& \frac{2\pi}{4+(\pi-2)^2} \frac{|B|}{ne} + \mathcal{O}(B^0) \label{1fluid_rho_highB} \\
	\lim_{B\rightarrow\infty}\rho_{xy} &=&  \frac{\pi(\pi-2)}{4+(\pi-2)^2}\frac{B}{nq} + \mathcal{O}(B^0)
	\label{1fluid_hall_highB}
\end{eqnarray}

Figure \ref{fig:1fluid}b) shows a series of $\rho(B)$ curves generated for such a 2D Fermi surface loosely based on a typical cuprate, including a scattering rate $\tau_0 = \hbar/k_BT$. The corresponding derivative curves are shown in Fig.~\ref{fig:1fluid}c). Eqs.~\eqref{rho_xx_1lfuid_low} and \eqref{rho_xy_1lfuid_low} confirm that the Drude result is recovered in the zero-field limit. The low-field $B$-linear regime is not seen experimentally. As we show below, however, this low-field $B$-linearity is found to be unstable to shorting effects and smoothness of the boundary. In line with experiment, Eq.~\eqref{1fluid_rho_highB} shows that the slope of the high-field $B$-linear MR is independent of $\tau_0$ (and $m^*$) and therefore independent of temperature. Indeed, as it turns out, the high-field MR is determined solely by the boundary (set by geometry through $W = \pi/2$). As shown in Fig.~\ref{fig:1fluid}c), the form of the MR also scales with $B/T$. Thus, many aspects of the quadrature MR behavior are reproduced with the simple introduction of an infinite barrier and a Planckian scattering rate. We reiterate that no MR emerges for the Drude result; all MR is generated solely by the presence of the boundaries. 

The high-field $B$-linear MR regime is reached at a magnetic field $B^*$, defined by the field strength where a quantum can travel between boundaries. This emerges theoretically through a characteristic exponent $\exp(-W/|\omega_c|\tau_0)$ in Eqs.~\eqref{sigmaxx_analytical} and \eqref{sigmaxy_analytical}. The turnover scale $B^*$ depends on microscopic details and is given by 

\begin{equation}
    B^*=Wm^*/e\tau_0
    \label{Bstar}
\end{equation}

 The derivative plot in Fig.~\ref{fig:1fluid}c) shows that this scale corresponds to the field at which $B$-linearity is, to all intents and purposes, fully established. Experimentally, the relevant field scale is usually defined as either the turning point (the knee in the derivative) or the point of deviation from a quadratic MR at low fields. The crossover is therefore typically assigned to significantly lower values than the intrinsic field scale of the problem as defined here. We also note that $B$-linearity can be observed at considerably lower fields than those at which quantum oscillations would be expected. Firstly, since typical materials satisfy or are close to $C_3$, $C_4$ or $C_6$ symmetry, multiple boundaries are expected on a Fermi surface, considerably lowering the relevant field scale compared to the full orbits required for quantum oscillations. Secondly, once quanta travel between boundaries, the derivative $d\rho/dB$ reaches $B$-linearity as shown in Fig.~\ref{fig:1fluid}c). These two effects allow for the onset of $B$-linear MR in the regime $\omega_c\tau_0<1$ and does not require $\omega_c\tau_0>2\pi$, a necessary criterion for a large number of explanations for $B$-linear MR \cite{Feng2019}.

\begin{figure}[hbtp]
	\includegraphics[width=\columnwidth]{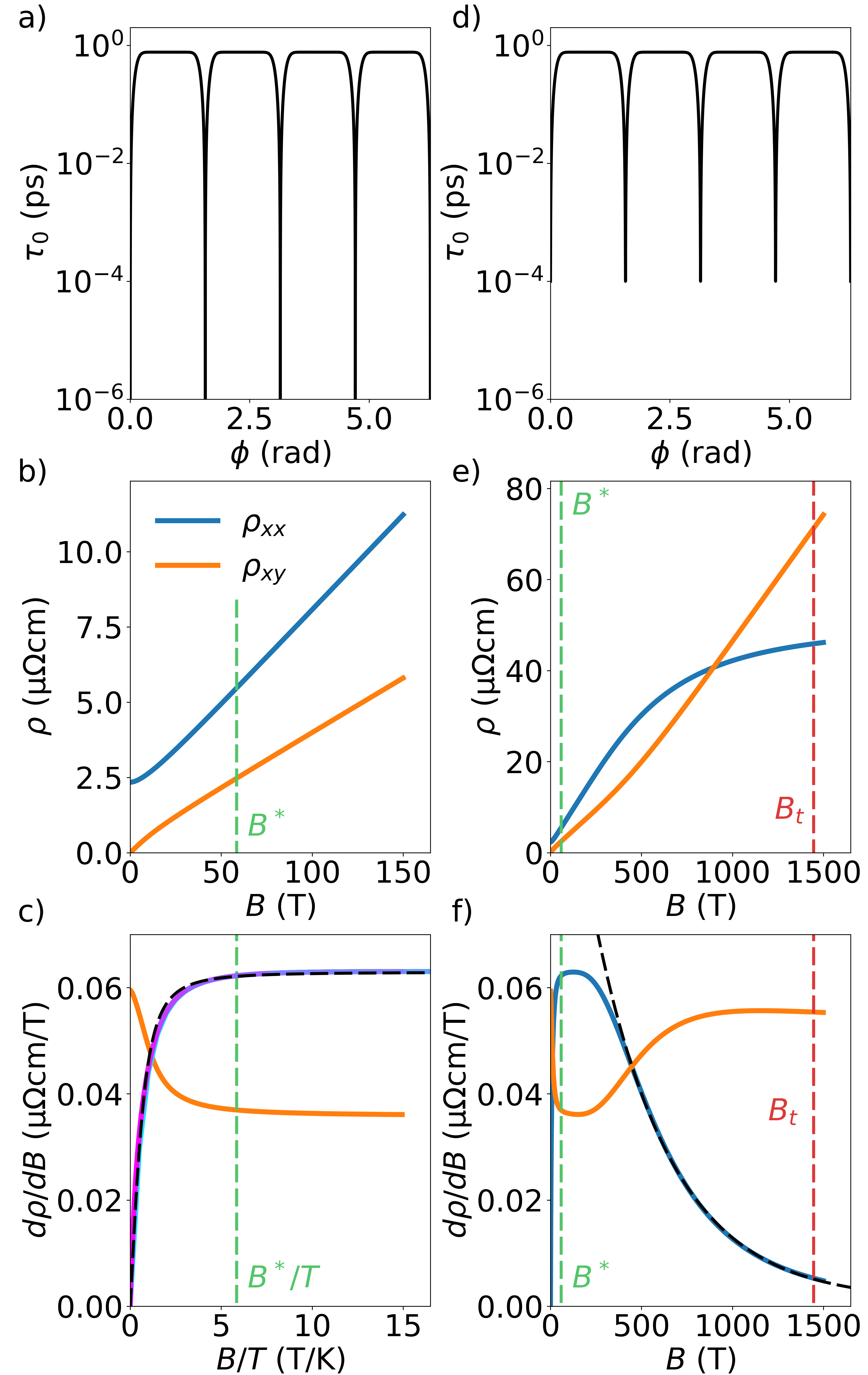}
	\caption{Lorentzian boundary model. a) Scattering rate around the Fermi surface. The hot spots have a Lorentzian shape with characteristic half-width of 0.1 rad. The cold scattering rate is given by $\tau_0=\hbar/k_BT$ at $T=10$ K. b) The resistivity and Hall effect using $k_F=7e9$ m$^{-1}$, $c=6.5$ $\mathrm{\AA}$ and $m^*=5m_0$, again based loosely on single-layer cuprates. The field scale $B^*=58$ T for impeded orbital motion is indicated in green across panels. c) The derivative shows QLMR behavior consistent with the quadrature form with a reduced field scale of 0.93 T/K as well as a linear high-field Hall effect. Varying the temperature shows a perfect scaling collapse in $B/T$, shown here for the temperature range 1 K (cyan) to 200 K (purple). d) Second model with reduced depth of the hot spots to illustrate the effect of penetration through the barrier and the irrelevance of the exact height of the scattering peak for the $B$-linear slope at intermediate field strengths. e) The corresponding resistivity to extremely high field shows saturation occurs when magnetic breakdown manifests. Shared with panel f), the characteristic tunneling field $B_t\sim 1500$ T is indicated in red (See Appendix \ref{appendix:generalized} for a derivation). f) d$\rho_{xx}$/d$B$ shows a low-field quadrature form with the same characteristics as in panel c). At high field, the barrier is penetrable and the MR saturates. The dashed line is an exponential fit with characteristic scale $\sim 400$ T above which the conventional Hall slope recovers and the MR saturates. }
	\label{fig:Lorentz}
\end{figure}

The obtained high-field regime is entirely unexpected from a traditional perspective. Again, Drude conductivity shows zero MR, while in the conventional Boltzmann scenario, anisotropy is eventually washed out leading to MR saturation at the highest fields. Here, without breaking the boundaries, the MR does not saturate and is non-accidental, as shown by the infinite field limit in Eq.~\eqref{1fluid_rho_highB}.

For the $\delta$-function barrier considered above, the low-field MR also contains a residual linear component that is not seen experimentally. Indeed, as we now show, this particular feature of the minimal model is found to be unstable to the introduction of arbitrary smoothness in the boundary. We emphasize that low-field $B$-linearity is an artifact of singular boundaries and not a central part of the phenomenology of impeded cyclotron motion. Following Ref.~\cite{Koshelev2016}, we model a smooth boundary by incorporating a locally enhanced scattering rate of Lorentzian shape on top of an isotropic background scattering rate. A similar smoothness of singularities was adopted in Refs.~\cite{Feng2019, Grissonnanche2020}. For these computations, we use the SCTIF formalism of Eq.~\eqref{sigma_general_no_temperature} to account for the continuously changing scattering rate around the Fermi surface. We show in panels a)-c) of Fig.~\ref{fig:Lorentz} that this scenario leads to a suppression of the linear low-field MR and a form that is highly consistent with the quadrature expression given in Eq.~\eqref{quadrature} (and shown as a dashed line in Fig.~\ref{fig:Lorentz}c)). In effect, smoothness of the boundary enables quanta to travel slightly beyond the impedance and prevents macroscopic charge from terminating at infinitesimal magnetic field, thereby shifting the influence of the boundary to finite fields and suppressing the initial $B$-linearity. One can interpret these findings as a gradual (as opposed to discontinuous) suppression of the effective scattering rate $\tau(B)$ near the boundary with increasing magnetic field. At the same time, the high-field $B$-linearity is dominated by the inter-boundary distance and is largely unaffected as the boundaries themselves remain impenetrable. We therefore find that even a minor softening of the boundary leads to a recovery of the quadrature MR. 

In a real material, of course, hot spots have non-zero scattering lifetime \cite{Koshelev2016, Koshelev2013}, van Hove singularities host a finite density of states \cite{Grissonnanche2020} and even true gaps can experience magnetic breakdown \cite{Naito1982}. In all cases, the probability to tunnel through the boundary scales as $\exp(-B_t/B)$ where $B_t$ corresponds to a second field scale at which the MR saturates exponentially in accordance with a diminishing anisotropy. This situation is illustrated in panels d)-f)  of Fig.~\ref{fig:Lorentz} for a slightly modified parameterization in which the depth of the scattering time at the hot-spots is reduced. In all cases, we find that linear MR is achieved at $B=B^*$, when quanta can traverse (on average) the entire region between boundaries, and is broken at $B=B_t$ when the boundaries themselves are penetrated, at which point the MR asymptotically approaches saturation. Analytically for a hot spot, this \lq breakdown' field can be understood as the probability of a quantum maintaining velocity-velocity correlation having reached a boundary and can be arbitrarily high. This contrasts with Ref.~\cite{Naito1982}, where $B$-linearity can only be reached through magnetic breakdown, whereas here $B$-linearity persists as long as breakdown does not occur. 

In those materials that have been shown to exhibit quadrature scaling, no violation of linearity has been observed to the highest $B/T$ values attained thus far: up to $B/T=35$ T/K in electron-doped cuprates with an estimated turnover scale of 2-3 T/K \cite{Sarkar2019} and up to 25 T/K in hole-doped cuprates with a turnover scale of 0.2 T/K \cite{Ayres2021}. This unwavering $B$-linearity up to the highest $B/T$ is one of the hallmarks of quadrature scaling. Intriguingly, SCTIF analysis of recent angle-dependent $c$-axis MR measurements on Nd-LSCO \cite{Grissonnanche2020} led to a very similar parameterization to the one shown in Fig.~\ref{fig:Lorentz}d)-f), a parameterization that also generates a $B$-linear in-plane MR extending well beyond the experimentally available field scale.

We now turn to consider the influence of impeded cyclotron motion on the Hall response. The geometrical factors in Eq.~\eqref{1fluid_hall_highB} lead to a marked reduction in the Hall conductivity compared to the Drude result. For a sector of general size, we derive in Appendix \ref{appendix:generalized} that $\sigma_{xy}$ is renormalized by a factor $1-\sin(W)/W$ in the high-field limit. This implies that in cases where less than half of the charge is in non-quasiparticle states and separated across four sectors ($W<\pi/4$), the Hall response can be suppressed by up to an order of magnitude. This geometrical factor may justify the approximation $\sigma_{xy}=0$ for the incoherent sector in overdoped cuprates used in Ref.~\cite{Culo2021}. Thus, we find that boundaries on the sector could simultaneously explain quadrature MR scaling and a suppression of $\sigma_{xy}$. 

\begin{figure}[ht!]
	\footnotesize
	\includegraphics[width=\columnwidth]{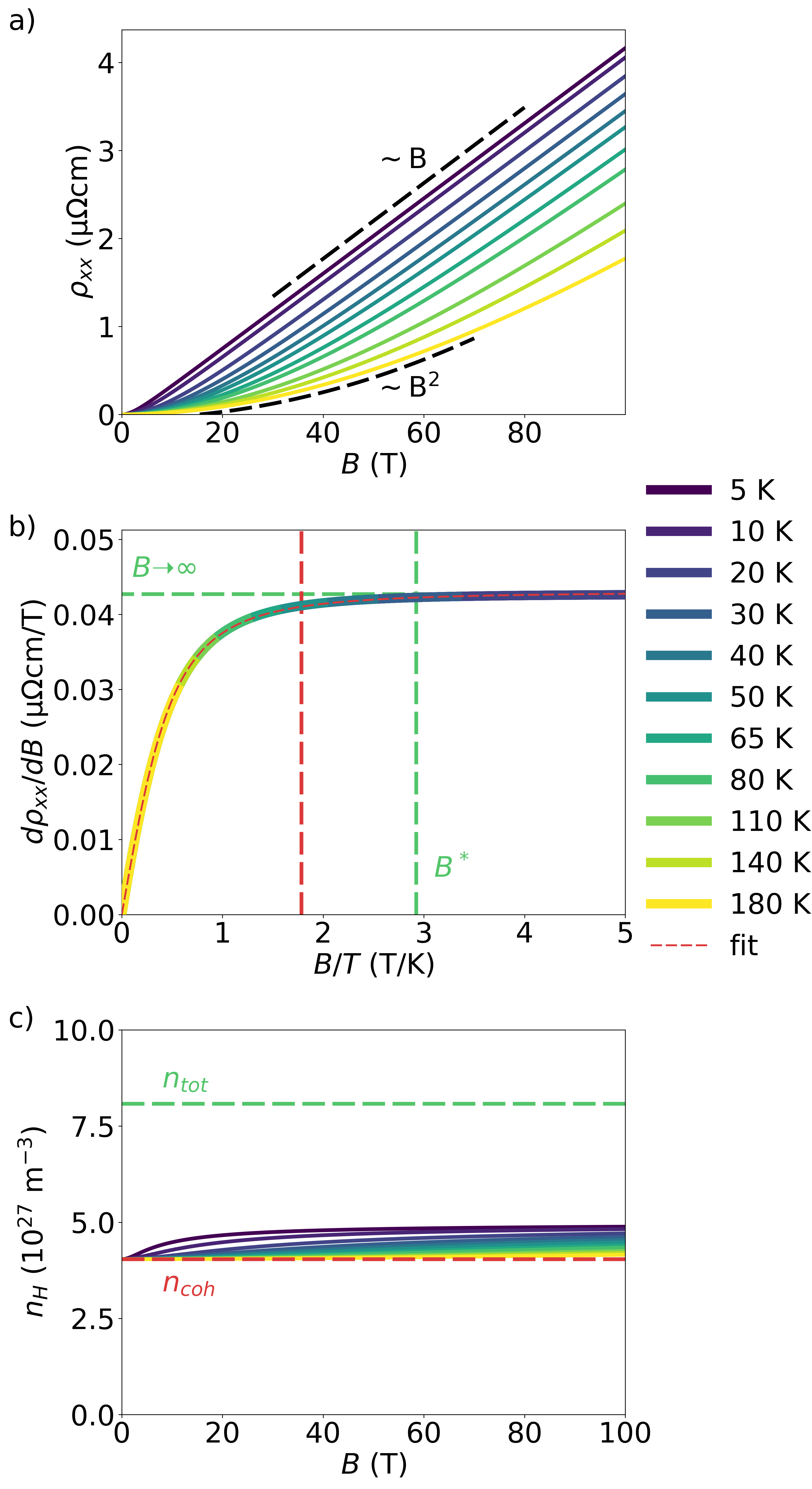}
	\caption{Impact of the introduction of multiple sectors on the QLMR. a) Resistivity with the same parameterization as in Fig.~\ref{fig:1fluid}, but with eight sectors. Four sectors are unbounded and host a 10000 times smaller scattering rate to represent the extreme shorting limit. Dashed lines indicate the dominant power law in the high- and low-$T$ limit. b) The derivative shows a scaling collapse. The dashed red lines are a fit to Eq.~\eqref{quadrature} and the associated field scale. The green vertical (horizontal) line represents the characteristic field scale $B^*$ (the theoretical saturation slope in the high-field limit), respectively. c) The corresponding Hall effect shows that the effective Hall number $n_{\rm H}$ is strongly reduced. Though the Hall effect remains $B$-linear, $n_{\rm H}$ is closer to $n_{coh}$, the number of coherent carriers (red dashed line), than $n_{tot}$ (green dashed line). }
	\label{fig:2fluid} 
\end{figure}

The similarity between the obtained high-field linear slope of $\rho_{xx}$ and $\rho_{xy}$ in Eq.~\eqref{1fluid_rho_highB} and~\eqref{1fluid_hall_highB} should also be noted. The high-field MR slope is found to be directly given by the carrier density (multiplied by a geometrical factor). There is no dependence upon $\tau_0$ (and therefore nor upon $T$), nor upon the effective mass (and therefore nor upon the energy dispersion or typical correlation effects). Intuitively, the $\tau_0$ independence emerges because only scattering on the boundary is relevant in the high-field limit. The independence of $m^*$ emerges because, although heavier carriers are harder to move, this effect is cancelled by the ability to survive longer until they encounter a boundary. The independence of a multitude of normally complicating factors suggests an insensitivity to material-specific parameters that reconciles the experimental fact that the magnitude of the high-field slope is highly robust as well as $T$-independent.

\section{Shorting effects}
\label{sec:shorting}

For the model presented thus far to be universal, it has to be robust to a number of generalizations. We discuss general $C_N$ symmetry in Appendix A. Here, we focus on shorting effects, which relate directly to the robustness of the results. To this end, we consider a two-fluid scenario whereby the Brillouin zone hosts two disparate Fermi pockets, one without borders (the conventional part) and one containing distinct bounded and unbounded sectors (the \lq strange' part). This may be relevant to multi-band systems in which boundaries only reside on a single sheet, or if a subset of quanta are, for whatever reason, able to traverse a boundary unimpeded. In the latter case, one can define two sectors on a single Fermi sheet. In either case, the conductivity $\sigma_c$ of the conventional sector is modelled within a Drude framework and is assumed to couple as a parallel resistor to the less conductive strange or bound contribution $\sigma_b$. In the limit $\sigma_{c} >> \sigma_{b}$, we expect shorting effects to suppress the quadrature MR, but as shown in Fig.~\ref{fig:2fluid}a), this is not observed. In fact, the derivative of the MR is unchanged, i.e. the result strongly resembles series MR coupling, despite the model in question being strictly parallel. 

Moreover, in the low-field limit, we find that the $B$-linear component -- already shown in Fig.~\ref{fig:1fluid}c) to vanish in the presence of smooth boundaries -- is also unstable to shorting effects. The key is in the inversion of the conductivity tensor to obtain the resistivity tensor (here we consider $\rho_{xx}$). In full:

\begin{eqnarray}
	\rho_{xx} = \frac{\sigma_{xx,b}}{(\sigma_{xx,b} + \sigma_{xx,c})^2+(\sigma_{xy,b}
	+ \sigma_{xy,c})^2} \nonumber\\ + \frac{\sigma_{xx,c}}{(\sigma_{xx,b}+\sigma_{xx,c})^2+(\sigma_{xy,b}+\sigma_{xy,c})^2}
	\label{shorting}
\end{eqnarray}

For $\sigma_{xx,b}<<\sigma_{xx,c}$, we find that the leading contribution of the bounded sector to the MR in the low-field limit is through $\sigma_{xx,b}/\sigma_{xx,c}^2$. This term is shorted out, thereby explaining the instability of the low-field $B$-linear component to shorting.

The more interesting scenario, however, is the high-field regime. Surprisingly, the high-field $B$-linear MR is found to be robust no matter how much more conductive $\sigma_{c}$ becomes. The underlying reason for this is that in the high-field limit $\sigma_{xx,c}\sim B^{-2}$ fades away much faster than $\sigma_{xx,b}\sim B^{-1}$ as well as both $\sigma_{xy}\sim B^{-1}$ components. (In Drude theory, the condition $\sigma_{xx,c}<<\sigma_{xy,c}$ is satisfied when $\omega_c\tau_0>1$.) The dominant $\sigma\sim B^{-1}$ scaling leads to robust $B$-linearity. Thus, we find that the low-field regime is dominated by $\sigma_{xx,c}$, whereas the high-field regime is dominated by the remaining three components.

The first term in Eq.~\eqref{shorting} is responsible for quadrature scaling while the second term defines the zero-field limit through the term $1/\sigma_{xx,c}$ and represents approximately conventional MR. Strikingly, approximate series coupling emerges in the high-field limit between the two components despite strictly parallel coupling underlying the model. For the Hall effect in the strongly shorted regime, we show in Fig.~\ref{fig:2fluid}c) that the Hall contribution is dominated by the coherent sector. Mathematically, the incoherent contribution is suppressed because $\sigma_{xy}$ is diminished by geometrical factors following Eq.~\eqref{sigmaxy_high_field_alpha}. 

The shorting argument presented above is extremely general, though in the intermediate regime where the difference between $\sigma_{xx,c}$ and $\sigma_{xx,b}$ become less severe, the field derivative $d\rho_{xx}/d\mu_0H$ is observed to go through a maximum before reaching the high-field plateau. Note that this is a rule of thumb and that a generalized treatment of this regime remains an open question and the relevance of these deviations are likely to depend on the situation in specific materials given the number of degrees of freedom involved. Nevertheless, the general result is that the high-field response is determined by $\sigma_{xx,b}$, $\sigma_{xy,b}$ and $\sigma_{xy,c}$, which are all defined by carrier density, carrier sign and geometry alone. By contrast, the zero-field resistivity is defined solely by $\sigma_{xx,c}$ and is dependent on both $\tau$ and $m^*$. As a result of this decoupling between the zero- and high-field limits, an effective series coupling between the zero-field resistivity and the MR emerges from fundamentally parallel coupling. We note that such series coupling between quadrature and conventional MR may have been observed in the iron chalcogenide family FeSe$_{1-x}$S$_x$ across its putative nematic QCP \cite{Licciardello2019PRR}. 
\section{Impurity scattering}
\label{sec:disorder}

The lack of a residual resistivity component in the MR scaling in strange metals \cite{Hayes2016, Maksimovic2020, Ayres2021} is one of the most striking and unusual features of their magnetotransport. Residual resistivity is usually incorporated via Matthiessen's rule, which for a Planckian metal can be expressed as $1/\tau_0=\alpha k_B(T+T_0)$, where $T_0$ originates from elastic impurity scattering and acts to reduce the velocity-velocity correlation decay time relative to the actual lifetime \cite{Hartnoll2015}. Experimentally, $T_0$ can be as high as 200 K \cite{Ayres2021}, making the distinction between quadrature $B/T$ scaling and conventional Kohler scaling through $\omega_c\tau_0\sim B/(T+T_0)$ unmistakable. 

In cuprates, the need to incorporate two lifetimes into the magnetotransport emerged early on following the discovery of a distinct $T$-dependence of the inverse Hall angle $\cot\theta_{\rm H}=\rho_{xx}/\rho_{xy}=a T^2$ + $b$ \cite{Chien1991lifetime, Chien1991HallT}. Original interpretations for such behavior included the spin-charge separation picture of Anderson \cite{Anderson1991} and $T$-dependent anisotropic scattering \cite{Carrington1992}. The coefficient $b$ was initially linked to the residual resistivity. Later MR measurements suggested that $\Delta\rho/\rho(0) \propto$ tan$^2\theta_{\rm H}$ \cite{Ando2002, Harris1995}, implying that $b$ was also a relevant parameter in the MR response. In cleaner cuprates, however, $\Delta\rho/\rho(0)$ was found to follow a pure power law \cite{Ando2002} or an adjusted form of Kohler scaling whereby plots of $\Delta\rho/(\rho(0)-\rho_0)$ as a function of $B/(\rho(0)-\rho_0)$ collapse onto a single curve \cite{Chan2014}, suggested a possible disconnect between the MR and the Hall angle. The origin of this behavior is still unknown.

The current model invokes a contribution to the conductivity which is activated by the application of a magnetic field and returns a $B$-linear slope of the resistivity that is independent of $\tau_0$, as highlighted in Eq.~\eqref{1fluid_rho_highB}. It is thus not immediately obvious that Matthiessen's rule will hold in this case since $\tau$ is $B$-dependent despite the underlying scattering rate on the Fermi surface being unchanged. Usually, Fermi's golden rule can explain the robustness of Matthiessen's rule for a wide variety of scattering types, but it is not immediately applicable. Not only is the boundary scattering induced by a magnetic field without changing any underlying scattering probabilities, impeded cyclotron motion also terminates carriers after a significant time period $t_{bound}$, whereas Fermi's golden rule concerns scattering over an infinitesimal time window.

In order to confront this problem, we have modeled impurity scattering for the simplest case using Boltzmann theory and isotropic impurity scattering, as described in Appendix \ref{appendix:disorder}. Incorporating this impurity scattering, we find that Kohler's rule is in fact recovered. We thus necessarily conclude that the reason why the residual resistivity is absent in the quadrature MR scaling lies beyond conventional Boltzmann theory.

\section{Discussion and Outlook}
\label{sec:discussion}

We begin this section by highlighting the key differences between impeded cyclotron motion and comparable alternative models for QLMR involving square Fermi surfaces \cite{Pippard1989}, hot spots \cite{Koshelev2016} and magnetic breakdown \cite{Naito1982}. We proceed by considering other possible candidates for the Fermi surface boundary before concluding with a brief discussion of QLMR in the high-$T_c$ cuprates, the archetypal strange metal \cite{Hussey2018}.

It has long been known that Fermi surfaces with sharp corners or turning points can give rise to $B$-linear MR. Even at small values of $\omega_c\tau_0$, charge may be forced to turn through a large angle ($\pi/2$ in the tetragonal case). Since the MR is driven by the mean-free-path orbit \cite{Harris1995}, these sharp corners will dominate the MR and since the turning angle is fixed, i.e. independent of field, the MR acquires a $B$-linear dependence. The sharper the turning point, the lower the value of $\omega_c\tau_0$ at which this $B$-linearity emerges. Conversely, a softening of the sharp corners (e.g., from a square to a rounded square) results in quadratic behaviour at low fields and hence QLMR. The same mechanism has been utilized in other models that invoke turning points \cite{Koshelev2013} and was recently proposed in favor of a myriad of alternatives requiring $\omega_c\tau_0>>1$ in CDW materials \cite{Feng2019}. In these scenarios, the scattering rate crucially remains unchanged in the vicinity of the turning point. 

Experimentally, it is doubtful whether such features are sufficiently general to be a universal explanation of QLMR. The Fermi surfaces of NbSe$_2$ \cite{Borisenko2009, Flicker2016}, pnictides \cite{Coldea2017, Shishido2010} and certain hole-doped cuprates \cite{Plate2005, Hussey2003, Vignolle2008}, for example, do not appear to contain such abrupt turning points. Only inside a phase of finite $Q$-nested order can such turning points emerge, though it is likely that these would also constitute strong scattering centers. Secondly, and perhaps more importantly, even for the perfectly square Fermi surface of Ref.~\cite{Pippard1989}, the $B$-linear MR turns out to be accidental, as shown in Appendix \ref{appendix:square_fs}. Quadrature MR, on the other hand, exhibits non-accidental $B$-linearity. Moreover, as shown here, impeded orbital motion can occur even on an isotropic Fermi surface with an anisotropic scattering rate. Finally, the field scale $B^*$ for impeded orbital motion is determined by the inter-boundary distance $W$, rather than the sharpness of the turning points. As a consequence, $B^*$ cannot be made arbitrarily small. Clearly, there are a number fundamental differences between the two pictures.

Many strange metals are located in the vicinity of a QCP of magnetic origin. The associated non-zero $\vec{Q}$ fluctuations can give rise to hot spots on the Fermi surface as well as non-Fermi-liquid $T$-linear resistivity. Hence, both ingredients necessary for the realization of QLMR with $B/T$ scaling are present in such systems. Our model is distinct from the previous treatment of quadrature scaling in Ba122 \cite{Koshelev2016, Koshelev2013, Maksimovic2020} -- that attributes QLMR to the intrinsic response of isolated hot spots beyond the relaxation time approximation -- in that here, the $B$-linear MR is tied to interactions between hot spots and the intrinsic response is shorted out \cite{Rosch1999}.

This difference manifests itself in a number of ways. Firstly, the intrinsic hot-spot model predicts $\rho_{xy} \propto B^2$ \cite{Koshelev2016}, whereas the current model correctly reproduces the observed $B$-linear Hall response. Secondly, the role of disorder in the temperature scaling is distinct for each model. In Ref.~\cite{Maksimovic2020}, impurity scattering must remain negligible compared to the scattering rate near the hot spot or the hyperbolic scaling breaks down. At the highest $B/T$, Kohler scaling is expected, but no deviations from $B$-linearity (let alone recovery of Kohler scaling) have ever been observed experimentally. 

Another model that is conceptually similar to ours was proposed in Ref.~\cite{Naito1982} to explain the observation of $B$-linear MR in NbSe$_2$. There, the $B$-linearity was claimed to originate from magnetic breakdown due to the fact that the rate at which CDW gaps on the Fermi surface are encountered scales with $\omega_c$. In our theorem, however, the effective anisotropy is washed out when magnetic breakdown occurs (see Fig.~\ref{fig:Lorentz}d)-f)) and $B$-linear MR can only occur in the absence of magnetic breakdown. This argument is further supported in Ref.~\cite{ Falicov1965} (and cited in \cite{Naito1982}) which shows MR saturation once perfect charge compensation is lost. 

As it turns out, our theorem may be equally applicable to materials like the dichalcogenides that undergo CDW formation. Partial gaps have been detected in TaSe$_2$ \cite{Borisenko2008} and NbSe$_2$ \cite{Borisenko2009} through ARPES and reproduced theoretically \cite{Flicker2016}, while $B$-linear MR is known to exist just above the superconducting dome \cite{Naito1982, Morris1972}. It is indeed likely that cyclotron motion is impeded as quantum oscillations from the pockets affected by the CDW have never been observed \cite{Graebner1976, Corcoran1994}. We thus speculate that impeded cyclotron motion is also responsible for the observation of QLMR in both these diselenides.

Another possible source of impeded orbital motion is proximity to a vHs. In the single-layer cuprate La$_{2-x}$Sr$_x$CuO$_4$ (LSCO), the Fermi surface undergoes a Lifshitz transition at a doping level $p \sim 0.20$ as the Fermi level crosses the vHs, resulting in enhanced density of states at the zone boundary and a pronounced scattering rate anisotropy. QLMR has now been reported in both LSCO \cite{GiraldoGallo2018} and Nd-doped LSCO \cite{Grissonnanche2020} at or near this doping level. Interestingly, the recent interlayer angle-dependent MR (ADMR) study in Ref.~\cite{Grissonnanche2020} also revealed that Nd-LSCO possesses the two key elements required for QLMR with $B/T$ scaling: a locus on the Fermi surface in which scattering is extremely large (thereby providing the boundary) and an isotropic Planckian scattering rate, thereby providing the quadrature scaling. Moreover, state-of-the-art optical conductivity measurements have indicated that the scattering rate in LSCO increases with increasing $B$ \cite{Post2020}, a possible direct manifestation of impeded cyclotron motion. The QLMR observed in electron doped cuprates \cite{Sarkar2019}, on the other hand, is unlikely to originate from proximity to a vHs. Instead, hot spots associated with the proximate AFM order is the most likely source of impeded orbital motion \cite{Armitage2002, Matsui2005}. 

Curiously, neither proximity to a vHs nor to the pseudogap endpoint $p^*$ can account for the realization of QLMR in two other hole-doped cuprates Tl$_2$Ba$_2$CuO$_{6+\delta}$ (Tl2201) and La/Pb-doped Bi$_2$Sr$_2$CuO$_{6+\delta}$ (Bi2201) \cite{Ayres2021}. ADMR \cite{Hussey2003}, ARPES \cite{Plate2005} and quantum oscillation \cite{Rourke2010} studies have all confirmed the presence of a full Fermi surface in overdoped Tl2201 ($T_c \sim$ 15-30 K), while a subsequent $T$-dependent ADMR study revealed that the mean-free-path becomes isotropic in the low-$T$ limit \cite{Abdel-Jawad2006}. Even though the ADMR parameterization could account for the $T$-dependent zero-field resistivity \cite{Abdel-Jawad2006}, it was later found to be inconsistent with the observation of in-plane QLMR and associated $B/T$ scaling \cite{Ayres2021}. Instead, both the QLMR \cite{Ayres2021} and the high-field Hall effect \cite{Putzke2019} were interpreted within a scenario in which the Fermi surface hosts distinct quasiparticle and non-quasiparticle sectors, the latter being postulated to form the superconducting condensate \cite{Culo2021}. The current work reinforces such a view by showing that if non-quasiparticle charge is bound to its sector on the Fermi surface, both QLMR and a reduced $\sigma_{xy}$ follow. 

While the origin of this boundary in overdoped Tl2201 and Bi2201 remains to be identified, the observation of $B/T$ scaling within the strange metal regime of both families suggests that it may be associated with Planckian dissipation itself. One important question to address here though is how effective borders can emerge in a system undergoing Planckian dissipation. Such $k$-space separation requires a non-quasiparticle quantum state. At the boundary of the strange sector, the quanta stop being good quantum numbers to the local Hamiltonian. This enables dissipation within the electronic system with energy uncertainty $E_f$ rather than $k_BT$, which allows through the uncertainty relation for fast enough dissipation to create a boundary. We remark that quanta would have to couple non-perturbatively to the dissipating degrees of freedom responsible for driving these states to the Planckian limit. This non-perturbative aspect is required for borders to emerge on the Fermi pocket because of simplical homology. Simply stated, this mathematical theorem shows no border can emerge on borders, meaning that no sector boundaries can emerge on the Fermi surface, which itself forms the border between occupied and unoccupied states. By mixing the degrees of freedom, we overcome this limitation. Intriguingly, this scenario can be interpreted as a loss of coherence at the edges of the strange metal sector. As we have shown, any impedance to cyclotron motion is sufficient to generate $B$-linear MR when solved for a general magnetic field. Suggestively, we point out a $1/T$ (rather than $1/B$) bound to the temporal coherence of quanta following the Planckian limit can similarly generate $T$-linear resistivity.

As mentioned in the Introduction, a further challenge to any successful theory of QLMR is to explain the dependence of the MR on field orientation and its contrasting behavior in the cuprates and pnictides. Specifically, while the magnitude of the quadrature MR in Ba122 scales with the field component orthogonal to the conducting plane \cite{Hayes2018}, the quadrature MR in cuprates turns out to be rather isotropic \cite{Ayres2021}. While the origin of this distinction is not known, one might speculate that in the presence of an in-plane field, orbital motion along $k_z$ in cuprates is also impeded while in pnictides, such a barrier is absent. The strong resistive anisotropy in cuprates -- orders of magnitude larger than in the pnictides --  may be one indication that some form of tunneling barrier along $k_z$ does indeed exist in the former. Within the isotropic 2D limit, the current theory predicts the resulting characteristic turnover scales to be related through $B^*_{c}/B^*_{ab}=2\pi/Wk_Fc$, where the subscript indicates the magnetic field direction. This ratio is of order unity, as observed \cite{Ayres2021}.

Arguably, the strongest evidence against impeded cyclotron motion being the origin of QLMR would be the observation of Shubnikov-de Haas (SdH) oscillations from the same charge in a region of field and temperature where the MR is strictly $B$-linear. In multi-band materials, of course, QLMR and SdH oscillations can coexist on distinct Fermi pockets. In a single-band material, however, the impedance cannot be avoided. In this regard, single-layer cuprates offer a stringent test of the present theorem. Quantum oscillations have been observed in overdoped Tl2201 \cite{Vignolle2008} in a doping regime where QLMR has also been detected \cite{Ayres2021}. At first sight, this coincidence appears to invalidate the notion of impeded cyclotron motion. It is important to note, however, that quantum oscillations have only currently been detected via torque or interlayer transport \cite{Vignolle2008, Rourke2010}, both of which can be argued to be single-particle, not particle-particle, probes \cite{Moses1999, Sandemann2001}. For in-plane transport, on the other hand, vertex corrections could well play a key role, even in the formation of the impedance itself. Thus, an acid test of this theorem would be a simultaneous search for SdH oscillations in $\rho_{zz}(B)$ and in $\rho_{xx}(B)$ (e.g. using suitable microfabrication techniques). Irrespective of the role of vertex corrections, and despite demonstrations to the contrary \cite{Chien1991HallT, Ayres2021}, the current findings do appear to support the notion that Boltzmann transport theory can be applied to strange and other strongly correlated metals. 

The present work set out to address the question whether or not there is a \textit{universal} explanation for the phenomenon of QLMR in correlated metals. It was shown that
impeded cyclotron motion can generate the quadrature form including scaling and non-accidental and non-saturating $H$-linear MR with sufficient universality. Putative origins for such impeded orbital motion include van Hove singularities, hot spots or hot lines, partially gapped Fermi surfaces, Fermi surface sectors caused by either AFM or CDW correlations and charge separation within a Fermi pocket. The resultant scaling is distinct from conventional Kohler's rule based on weak and smooth anisotropy yet remains orbital in nature. The central outstanding feature of quadrature phenomenology that still lies beyond Boltzmann theory is the apparent absence of elastic (impurity) scattering in the QLMR scaling. Future studies into this aspect of strange metals may well be the key to unlocking the mystery of their normal, i.e. non-superconducting, states.

\appendix

\section{Generalized sector size}
\label{appendix:generalized}

In this section, we generalize the simple 1-fluid scenario to $C_N$ symmetric systems, to other orientations of the impedance and to partially bounded Fermi surfaces. We will show that $N$ distinct mirror planes through the impedances (as guaranteed by $C_N$ symmetry) is sufficient to conserve the expected time and charge inversion symmetries. Furthermore, as claimed in the main text, for $N>2$ the angle degree of freedom for the location of the boundary on the Fermi surface (e.g. diagonal or horizontal/vertical in the tetragonal case) is shown to be irrelevant for the total conductivity in the absence of further anisotropy. Finally, we find for a general sector size the suppression of the Hall conductivity, the $B$-linear slope and turnover scale $B^*$ as given in the main text.

We restrict ourselves throughout to 2D Fermi pockets, though we note that boundaries can equally be manifested in three dimensions, e.g. through hot lines as proposed in Ref.~\cite{Rosch1999}. In 3D, we anticipate much of the same physics, but depending on the field direction not all quanta will necessarily encounter a boundary resulting in open orbits. It is well known that open orbits heavily influence the MR \cite{Lifshitz1957} and such a case would deserve separate treatment. 

Mathematically, we consider an isolated sector centered at azimuthal angle $\alpha_0$ and with boundaries extending an angle $\alpha$ to either side. Adopting Eq.~\eqref{sigma_general_bounded} for electron-like quanta ($q<$0 thus $\omega_c<$0), we obtain:

\begin{align}
	\sigma_{xx} = & \frac{e^2}{2\pi\hbar c} \int_{\alpha_0-\alpha}^{\alpha_0+\alpha} d\phi \int_0^{(\alpha_0-\alpha-\phi)/\omega_c}dt 
	\nonumber \\ &\quad
	k_F(0) \frac{v_x(0)}{v_F(0)} v_x(-t)\exp(-t/\tau_0)
	\label{sxx_strange_quadrant_2fluid}
\end{align}

Solving these integrals for a perfect 2D Fermi surface with the same assumptions as in section \ref{sec:specific}, we find the rather complicated looking expressions shown below. 

\begin{align}
	\sigma_{xx,s} = & 
	    \frac{n_s e^2\tau_0}{m^*(1+\omega_c^2\tau_0^2)}
	\left(
	    1  + \frac{\cos(2\alpha_0)\sin(2\alpha)}{2\alpha} - 
	    \nonumber \right.\\&\left.
	    \frac{\omega_c\tau_0}{2\alpha}\sin(2\alpha_0)\sin(2\alpha) +
	    \nonumber \right.\\&\left.
	    \frac{1}{\alpha}\frac{\omega_c\tau_0}{1+\omega_c^2\tau_0^2}
	    \left(
	        \omega_c\tau_0\sin(\alpha_0-\alpha) - \cos(\alpha_0-\alpha)
	    \right)
	    \centerdot 
	    \right.\nonumber\\&\left.
	    \Big(
	        -\cos(\alpha_0-\alpha) -\omega_c\tau_0\sin(\alpha_0-\alpha)+
	        \nonumber\right.\\ &\left.
	        e^{2\alpha/\omega_c\tau_0}
	        \left(
	            \omega_c\tau_0\sin(\alpha+\alpha_0)+\cos(\alpha_0+\alpha)
            \right)
	    \Big)
	\right)
	\label{sigmaxx_2fluid_initial}
\end{align}

\begin{align}
	\sigma_{yy,s} = & 
	\frac{n_s e^2\tau_0}{m^*(1+\omega_c^2\tau_0^2)}
	\left(
	    1 - \frac{\cos(2\alpha_0)\sin(2\alpha)}{2\alpha} +
	    \nonumber \right.\\&\left.
    	\frac{\omega_c\tau_0}{2\alpha}\sin(2\alpha_0)\sin(2\alpha)
    	+ \frac{1}{\alpha}\frac{\omega_c\tau_0}{1+\omega_c^2\tau_0^2} \centerdot
    	\nonumber \right.\\&\left.
    	\left(
    	    \omega_c\tau_0\cos(\alpha_0-\alpha) + \sin(\alpha_0-\alpha)
    	\right)
    	\centerdot
    	\nonumber \right.\\&\left.
    	\Big(
    	    \sin(\alpha_0-\alpha)
    	    -\omega_c\tau_0\cos(\alpha_0-\alpha)+
    	    \nonumber \right.\\&\left.
    	    e^{2\alpha/\omega_c\tau_0}
    	    \left(
    	        \omega_c\tau_0\cos(\alpha+\alpha_0)-\sin(\alpha_0+\alpha)
	        \right)
    	\Big)
	\right)
	\label{sigmayy_2fluid_initial} 
\end{align}

\begin{align}
	\sigma_{xy,s} = & \frac{n_se^2\tau_0}{m^*(1+\omega_c^2\tau_0^2)}
	\left(
    	\omega_c\tau_0 
    	- \frac{\sin(2\alpha_0)\sin(2\alpha)}{2\alpha} -
    	\nonumber\right.\\&\left.
    	\frac{\omega_c\tau_0}{2\alpha}\cos(2\alpha_0)\sin(2\alpha)
    	+\frac{1}{\alpha}\frac{\omega_c\tau_0}{(1+\omega_c^2\tau_0^2)} 
    	\nonumber \right.\\&\left.
    	\left(
    	    \omega_c\tau_0\sin(\alpha_0-\alpha) -\cos(\alpha_0-\alpha)
	    \right) 
	    \centerdot
	    \nonumber\right.\\&\left.
    	\Big(
        	\sin(\alpha_0-\alpha) 
        	-\omega_c\tau_0\cos(\alpha_0-\alpha) -
        	\nonumber \right.\\&\left. 
        	e^{2\alpha/\omega_c\tau_0}
        	\left(
        	    \sin(\alpha_0+\alpha) - \omega_c\tau_0\cos(\alpha_0+\alpha)
    	    \right)
    	\Big)
	\right)	
	\label{sigmaxy_2fluid_initial}
\end{align}
We repeat the procedure for holes, for which the upper time bound changes to $(\alpha_0+\alpha-\phi)/\omega_c$. These results are similar and not shown. We have introduced $n_s=n\alpha/\pi$ to note the charge contained within each sector.

The above equations can be simplified considerably when we consider $C_N$ symmetry as realized in most strange metals. We now proceed to show that all dependence on $\alpha_0$ disappears for $N>2$, though some modifications can exist in orthorhombic cases where $\sigma_{xx}\ne \sigma_{yy}$.

The total conductivity $\sigma_{xx}$ is the total of multiple sectors which, by symmetry, only differ in $\alpha_0$ value. $C_N$ symmetry informs that the vector sum of the center locations of the sector centers in $k$-space is the center of the pocket itself. Mathematically, this means $\sum_i \cos\alpha_{0,i}=0$ and $\sum_i \sin\alpha_{0,i}=0$. One can deduce through standard goniometry that Eqs.~\eqref{sigmaxx_2fluid_initial}-\eqref{sigmaxy_2fluid_initial} can be rewritten such that $\alpha_0$ only appears through $\sin(2\alpha_0)$ or $\cos(2\alpha_0)$ terms. Summing over all $N$ sectors in $C_N$ symmetry, these terms vanish if $N>2$. Consequently, $\sigma$ is independent of $\alpha_0$ and the orientation of the boundaries is irrelevant. This argument holds even if the bounded charge does not cover the entire Fermi surface.


Informed by these symmetry arguments, we aim to reconstruct the full conductivity rather than the contribution of a single sector to simplify the results. In the end, $N$ in $C_N$ only enters to obtain the total charge contained in the sectors and implicitly as $\alpha$ is limited to $\pi/N$. We end up with the following for the total conductivity, which is valid for both electrons and holes:

\begin{align} 
	\sigma_D :=& \frac{n_s N e^2\tau_0}{m^*(1+\omega_c^2\tau_0^2)} \\
	\sigma_{xx} =& 
	\left(
    	1 - \frac{1}{2\alpha}\frac{|\omega_c\tau_0|}{1+\omega_c^2\tau_0^2}
    	\Big(
        	\left(
        	    1 - \omega_c^2\tau_0^2
    	    \right)
    	    \centerdot
    	    \nonumber\right.\\&\left.
        	\left(
        	    1 - \cos(2\alpha)e^{-2\alpha/|\omega_c\tau_0|}
    	    \right) +
    	    \nonumber\right.\\&\left.
        	2|\omega_c\tau_0| \sin(2\alpha) e^{-2\alpha/|\omega_c\tau_0|}
    	\Big)
	\right) \sigma_D 
	\label{sigmaxx_2fluid}\\
	\sigma_{xy} =& 
	\left(
	    1 - \frac{1}{2\alpha} \frac{1}{1+\omega_c^2\tau_0^2} 
	    \Big(
	        2|\omega_c\tau_0|
    	    \centerdot
    	    \nonumber\right.\\&\left.
	        \left(
	            1-\cos(2\alpha)e^{-2\alpha/|\omega_c\tau_0|}
            \right) 
            - 
            \nonumber \right. \\ &\left.
            \left(
                1-\omega_c^2\tau_0^2
            \right)
            \sin(2\alpha) e^{-2\alpha / |\omega_c\tau_0|}
        \Big)
	\right) 
	\sigma_D\omega_c\tau_0
	\label{sigmaxy_2fluid}
\end{align}

Here, $\sigma_D$ is the Drude response, with $n_s$ the sector carrier density and $Nn_s$ the total charge considered. These closed form results underpin the quadrature form and allow for introspection. They also show interesting parallels between on- and off-diagonal components.  

We start by remarking that the field scale governing these contributions is disambiguously $|\omega_c\tau_0|=2\alpha$ from the exponent, which defines $B^*$. This corresponds to the notion that charge can travel between boundaries in expectation value given the zero-field scattering lifetime. The definition of $B^*$ can be made one step more general still by introducing additional anisotropy, in which case SCTIF replaces $\exp(-t/\tau_0)$ by $\exp\left(-\int \frac{dt}{\tau_0}\right)$ as in Eq.~\eqref{sigma_general_no_temperature}. If the time integral is subsequently (effectively) truncated at a border, the ultimately relevant field scale is when quanta travel between boundaries in expectation value given their zero field lifetime. In the most general form considered here, we thus refine the definition of $B^*$ to the following, where we note angle dependence explicitly for emphasis.

\begin{equation}
    B^* = B \int_{\text{bound A}}^{\text{bound B}} \frac{d\phi}{|\omega_c(\phi)\tau_0(\phi)|} = \frac{1}{e} \int_{\alpha_0-\alpha}^{\alpha_0+\alpha} \frac{m^*(\phi) d\phi}{\tau_0(\phi)}
\end{equation}

In practise, for the Lorentzian boundary scenario considered in Section \ref{sec:specific}, a slight uncertainty in the definition remains as the boundaries are smooth and have to be excluded from the integration, which is somewhat ambiguous. However, we find the maximum uncertainty to be at most a few percent.

Returning to Eqs.~\eqref{sigmaxx_2fluid} and \eqref{sigmaxy_2fluid}, the renormalisation compared to the Drude result is once again found to be insensitive to the sign of the charge, resulting in the standard particle-hole symmetries of $\sigma_{xx}$ and $\sigma_{xy}$. The terms in brackets would simplify to 1 if the borders were removed, i.e. when the upper bound of the time integral is set to infinity, as is the case in the $B\rightarrow 0$ limit. This highlights the fact that boundaries are the only non-Drude element present and do not affect the zero-field behavior. 

The Drude term emerges from the lower time bound in Eq.~\eqref{sxx_strange_quadrant_2fluid}, whereas the unconventional term introduced here is the result of a finite upper time bound, showing a clear separation between the two responses. As a result, parallel resistor coupling is guaranteed $\sigma=\sigma_{Drude}+\sigma_{bound}$ with $\sigma_{bound}=0$ in the absence of a magnetic field. This argument holds no matter the kind of microscopics leading to the manifestation of a border.

We further note that Eqs.~\eqref{sigmaxx_analytical} and \eqref{sigmaxy_analytical} are derived as a special case from Eqs.~\eqref{sigmaxx_2fluid} and \eqref{sigmaxy_2fluid} by invoking $N=4$, $\alpha=\pi/4$ and $n=4n_s$. Using a Laurent series and matrix inversion the following limits are obtained. We use $n$ here instead of $n_sN$ to denote the total bounded charge.

\begin{eqnarray}
	\lim_{B\rightarrow\infty} \sigma_{xx} &=& \frac{1-\cos(2\alpha)}{2\alpha} \frac{ne}{B} 
	\label{sigmaxx_high_field_alpha}\\
	\lim_{B\rightarrow\infty} \sigma_{xy} &=& \left(1 - \frac{\sin(2\alpha)}{2\alpha}\right)\frac{nq}{B}
	\label{sigmaxy_high_field_alpha}
\end{eqnarray}

In the case of 4 boundaries on the Fermi surface, the Hall conductivity was found to be reduced by 64 \% in Eq.~\eqref{1fluid_hall_highB}. For hexagonal symmetry, we find using the generalized expressions that the Hall conductivity reduces in the high-field limit by 83 \% compared to the standard Fermi-liquid value. With 4 sectors covering half the Fermi surface or less, which is a case not unlike those suggested in Ref.~\cite{Culo2021}, the Hall conductivity can be suppressed by a full order of magnitude. This validates the empirical approximation $\sigma_{xy}=0$ used in their analysis as a necessary consequence of the sectors introduced on the Fermi surface. Combined with quadrature MR in the resistivity response, we have thus found a deeper phenomenology to explain these findings.

For the resistivity, we find the following: 

\begin{eqnarray}
	\lim_{B\rightarrow\infty} \rho_{xx} &=& 
	\frac{2\alpha(1 - \cos(2\alpha))}{(1-\cos(2\alpha))^2 + (2\alpha - \sin(2\alpha))^2}
	\frac{B}{ne} \label{rhoxx_high_field_alpha}\\
	\lim_{B\rightarrow\infty} \rho_{xy} &=& 
	\frac{2\alpha - \sin(2\alpha)}{(1-\cos(2\alpha))^2 + (2\alpha - \sin(2\alpha))^2}
	\frac{B}{nq} \label{rhoxy_high_field_alpha}
\end{eqnarray}

which simplifies to Eq.~\eqref{1fluid_rho_highB} and \eqref{1fluid_hall_highB} when $\alpha=\pi/4$.
\section{Disorder Modelling}
\label{appendix:disorder}

If quadrature scaling is to be explained through cyclotron effects, one of the key questions to address is how the residual resistivity can be absent from the scaling. In this Appendix, we first outline why Matthiessen's rule is not immediately apparent for this type of MR. Subsequently, we present Monte Carlo simulations using the simplest isotropic impurity scattering model possible. As outlined in the main text, this modelling shows that Kohler scaling is {\it not} violated in the simplest case and thus the insensitivity to impurity scattering cannot be understood from such a picture.

In the most general sense, impeded cyclotron motion drives MR through the saturation of $\omega_c\tau(B)$. In combination with $\omega_c\sim B$, this gives rise to $\tau\sim 1/B$. Ultimately, the effective relaxation time is dependent on magnetic field, yet the anisotropic scattering rate on the Fermi surface is unchanged with the application of magnetic field. This scenario is difficult to describe using Matthiessen's rule as it assumes the scattering time and relaxation time are equal.  

Furthermore, boundary scattering emerges through an effective time truncation of the velocity-velocity correlation as shown in Eq.~\eqref{sigma_general_bounded}. Since this correlation is continuous in time, the first fundamental theorem of calculus is applicable and the result of the time integral may be written as $F(\infty)-F(0)$ for some unknown functional $F$. Standard resistivity computations only focus on the $F(0)$ term, as the upper bound is infinity and no velocity-velocity correlation can persist indefinitely. In other words, $F(\infty)=0$. However, by introducing an effective boundary, we have made the tail of the distribution important, which manifests through a finite $F(t_{bound})$. A parallel resistor scenario necessarily follows which cannot be understood from a Fermi's golden rule nor from a Matthiessen's rule treatment of the scattering rate. 

For simplicity, we consider here a scenario in which elastic impurity scattering is isotropic. As a result, velocity-velocity correlation vanishes in the macroscopic average after a scattering event and impurities manifest through an effective upper bound on the time integral, which in this case depends on the real-space position of the quantum under consideration and is isotropic in $k$-space. In the relaxation time approximation, it is invoked that the emergent effective decay rate is proportional to the surviving quanta at that time. This results in exponential decay $\exp(-t/\tau_0)$ of the correlation function. 

We then arrive at the central question: When a magnetic field is applied and both real-space impurity and $k$-space boundary scattering enter the upper time bound, does conventional Kohler's rule hold? If so, does Matthiessen's rule hold for the $\omega_c\tau_0$ scaling of the MR?

\subsection{Model for Monte Carlo simulations}

To answer these questions, we consider a patch of real space in 2D with an isotropic Fermi surface for which we perform the following integral.

\begin{equation}
	\sigma_{ij} = \frac{e^2}{4\pi^3 A} \int_{patch}d^2r \int_{FS}d^2k \int_0^{\tau_{min}} dt 
	\frac{v_i(0)v_j(-t)}{v_F(0)}  e^{-t/\tau_0}
\end{equation}

Here, $A$ is the area of the real-space patch and $\tau_{min}$ is the minimum time for either the first Fermi surface boundary to be encountered or the first impurity to be scattered from and depends on both $k$ and $r$, which we will omit for clarity. $\tau_0$ amounts to any additional scattering from Fermi's golden rule, such as phonons or magnons. To keep degrees of freedom minimal while introducing a temperature dependence, we consider $\tau_0=\hbar/k_BT$. Furthermore, we consider the case of 4 boundaries on the Fermi surface as in section \ref{sec:specific}. We take the real-space patch to be a square with sides $L$ (typically 0.2$\mu$m). Substituting the standard relation $\vec{v}_F=\hbar\vec{k}/m^*$, equating the $k_z$ integral to $2\pi/c$ and simplifying $dk_{\phi}$ to $k_Fd\phi$, we find:

\begin{align}
	\sigma_{xx} = 
	\frac{e^2k_F^2}{2\pi^2 m^* L^2 c}  
	&\int_0^L dx \int_0^L dy \int_0^{2\pi} d\phi\int_0^{\tau_{min}} dt 
	\nonumber\\
	& \cos\phi\cos(\phi + \omega_ct) \exp(-t/\tau_0) 
\end{align}

The time integral can be evaluated in closed form as follows, manifesting a distinct contribution from the $t=0$ and $t=\tau_{min}$ bounds. The $t=0$ contribution is the Drude contribution. 

\begin{align}
	\sigma_{xx} = &
	\frac{e^2k_F^2}{2\pi^2 m^* L^2 c} \frac{\tau_0}{1+(eB\tau_0/m^*)^2}
	\int_0^L dx \int_0^L dy \int_0^{2\pi} d\phi
	\nonumber \\
	&\cos\phi
	\left( 
	    \cos\phi - \cos(\phi + \omega_c\tau_{min})\exp(-\tau_{min}/\tau_0) +
	    \right. \nonumber \\ &\left. 
	    \omega_c\tau_0 \sin(\phi + \omega_c\tau_{min})\exp(-\tau_{min}/\tau_0)
	    - \omega_c\tau_0 \sin(\phi)
    \right)
	\label{MonteCarloFormula}
\end{align}

The remaining integrals have to be performed using a Monte Carlo simulation as they rely on the exact impurity distribution. For the numerical evaluation of the spatial integrals, we evenly distribute positions and use Simpson's rule, typically 255x255. For each point, we simulate an entire Fermi surface through the $\phi$ integral. For each quantum simulated in this manner, the goal is to determine $\tau_{min}$ in order to evaluate the integrand in Eq.~\eqref{MonteCarloFormula} and the corresponding expression for $\sigma_{xy}$.

Impurities are randomly generated using a uniform distribution and are taken to be circular with a radius of 1 nm, comparable to an in-plane lattice constant. Having generated the impurity environment, the next step is to develop an efficient algorithm to compute the first impurity intersect for a given starting position and angle in order to determine $t_{min}$. Because the magnetic field is critical to keep the $k$-space boundary relevant, the trajectory of the quantum has to be described using cyclotron motion. We already introduced an isotropic Fermi surface with isotropic effective mass defining the velocity. We now furthermore assume $m_c=m^*$ to obtain the cyclotron radius as follows.

\begin{equation}
	r_{cycl} = \frac{m^*v_F}{Be}
\end{equation}

However, the cyclotron orbit can drive quanta outside the $L^2$ patch of real space. We thus have to consider boundary conditions. Fixed boundary conditions would amount to terminating moving charge at the boundary of the region. Periodic boundary conditions amount to teleporting the charge to the opposite side of the square. This choice is highly influential on the results and neither option reflects a bulk material. 

Instead, we create a larger region containing the patch of interest. This larger region (typically $\ge 10$ times the size) is where impurities are uniformly distributed with a given density. Because of the nature of impurity scattering, it is expected that the length scale over which charge is scattered is exponentially distributed $\frac{1}{\ell}\exp(-r/\ell)$ for a particular mean free path $\ell$ that is non-trivially related to the impurity density and radius. Even when $\ell$ is known through simulation, the inherently statistical nature prevents a guarantee that none of the quanta escape and requires a larger area if more carriers are considered. To make sure the bulk environment is sufficiently large, we track the number of times a carrier escapes the environment and make sure this number is zero after all computations finish.

\subsection{Procedure for Monte Carlo simulations}

Computationally, the evaluation of Eq.~\eqref{MonteCarloFormula} primarily concerns the determination of $\tau_{min}(x,y,\phi)$. This scattering time amounts to the earliest encounter with either a $k$-space boundary or $r$-space impurity. The former is given by $\Delta\phi/|\omega_c|$, where $\Delta\phi$ is measured between the currently considered $\phi$ and the first impedance to be encountered, taking into account hole- or electron-like carriers. The latter is given by the time it takes to reach the first impurity encountered. Mathematically, this scenario simplifies to intersections between cyclotron orbits and impurities.

In a naive approach, we could find the first impurity intersect by testing every impurity. However, there are typically millions of impurities to test, which quickly makes the computation unfeasible.

In order to improve the situation, we adopt a spatial indexing approach. After the impurities have been uniformly distributed, they are divided into cells in a grid (typically a statistical average of 10 impurities per cell was found to be optimal). It is common that impurities overlap multiple grid cells, in which case they are added to each cell.

Given a quantum, its position can be mapped to a grid cell. Then, we proceed to test for intersections between the cyclotron orbit and all present impurities in that cell. If no impurity intersects with the cyclotron orbit, we move to the next cell along the cyclotron orbit. Critically, the first cell in which an intersection is found corresponds to the first intersection globally.

Through this construction, on average only 1.5 - 3.5 grid cells had to be traversed per quantum. Consequently, these computations only require looking up a tiny fraction of the millions of impurities present in the bulk environment, significantly lowering the computation time. Additionally, the contribution of every quantum can be independently computed and this allowed for further improvement of the computation time through parallel execution.

\begin{figure}[hbtp]
	\includegraphics[width=\columnwidth]{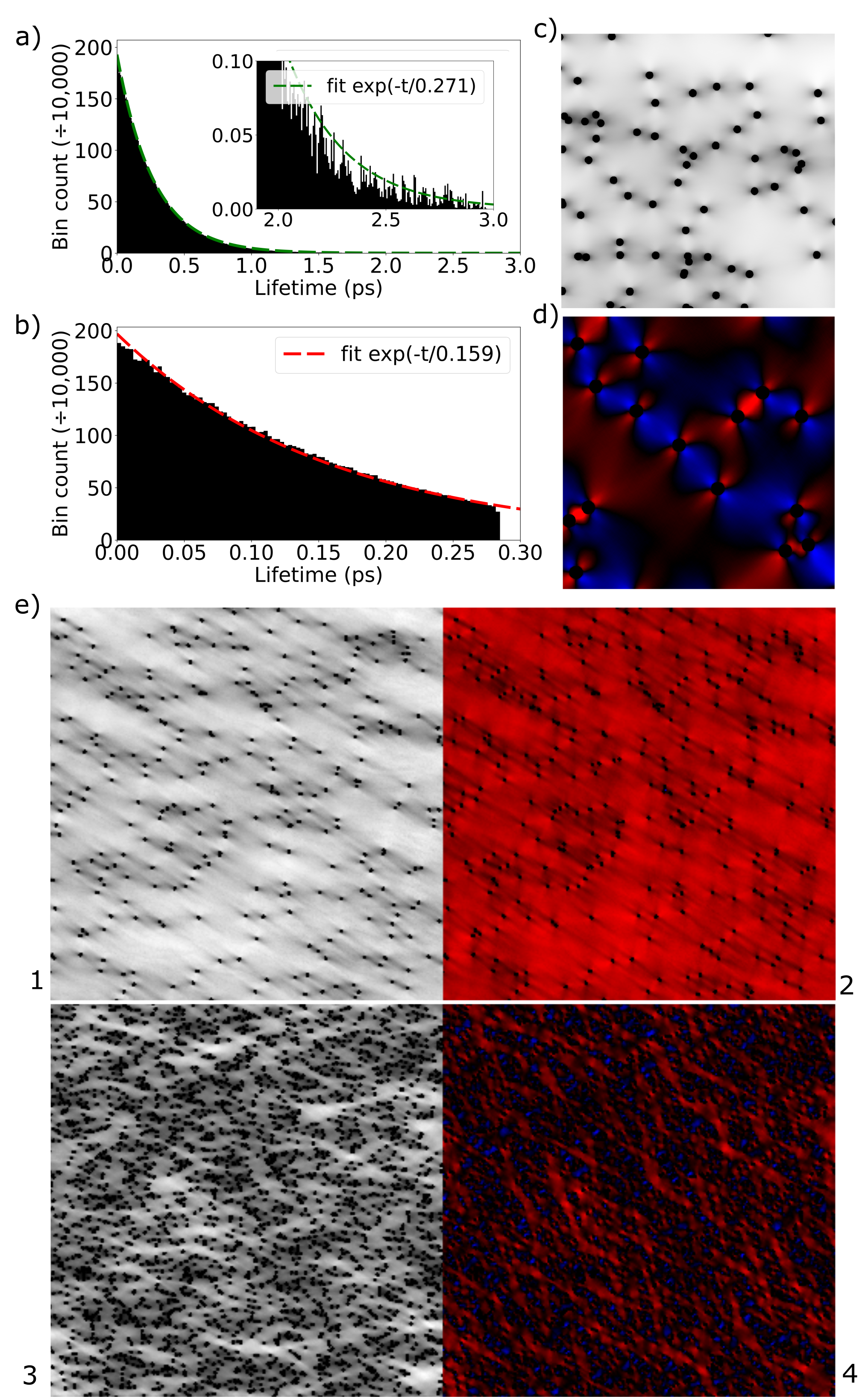}
	\caption{Examples from the disorder model. a) Coherent (unbounded) quanta follow the exponential decay from the relaxation time approximation within statistical accuracy. b) Incoherent (bounded) quanta do as well, but the velocity-velocity correlation function is terminated at a particular time where $\omega_ct$ matches the inter-boundary distance. c) Real-space image of $\sigma_{xx}$ in a limited real-space patch. Each pixel in this image represents a full Fermi surface integral (all velocity directions) in Eq.~\eqref{MonteCarloFormula}. White is high conductivity, whereas black is zero conductivity, which occurs inside impurities. $\sigma_{xx}$ is locally suppressed in the $x$ direction away from impurities. d) Similar to panel c) but for the Hall effect at zero magnetic field in a very limited real-space patch. Red and blue correspond to opposite sign contributions which turn black on approach to zero. This shows $\sigma_{xy}=0$ is locally broken, yet globally conserved. e) Real-space snapshots at highest magnetic field and lowest temperatures used in Figure \ref{fig:disorder}d-f) (panels 1,2) and Figure \ref{fig:disorder}a-c) (panels 3,4). The panels show both $\sigma_{xx}$ (panels 1,3) and $\sigma_{xy}$ (panels 2,4). These snapshots are all 0.2 $\mu$m in size. In panels d) and e), $\sigma_{xy}$ is red when positive or hole-like and blue when negative or electron-like. }
	\label{fig:disorder_model}
\end{figure}

\begin{figure}[hbtp]
    \centering
    \includegraphics[width=\columnwidth]{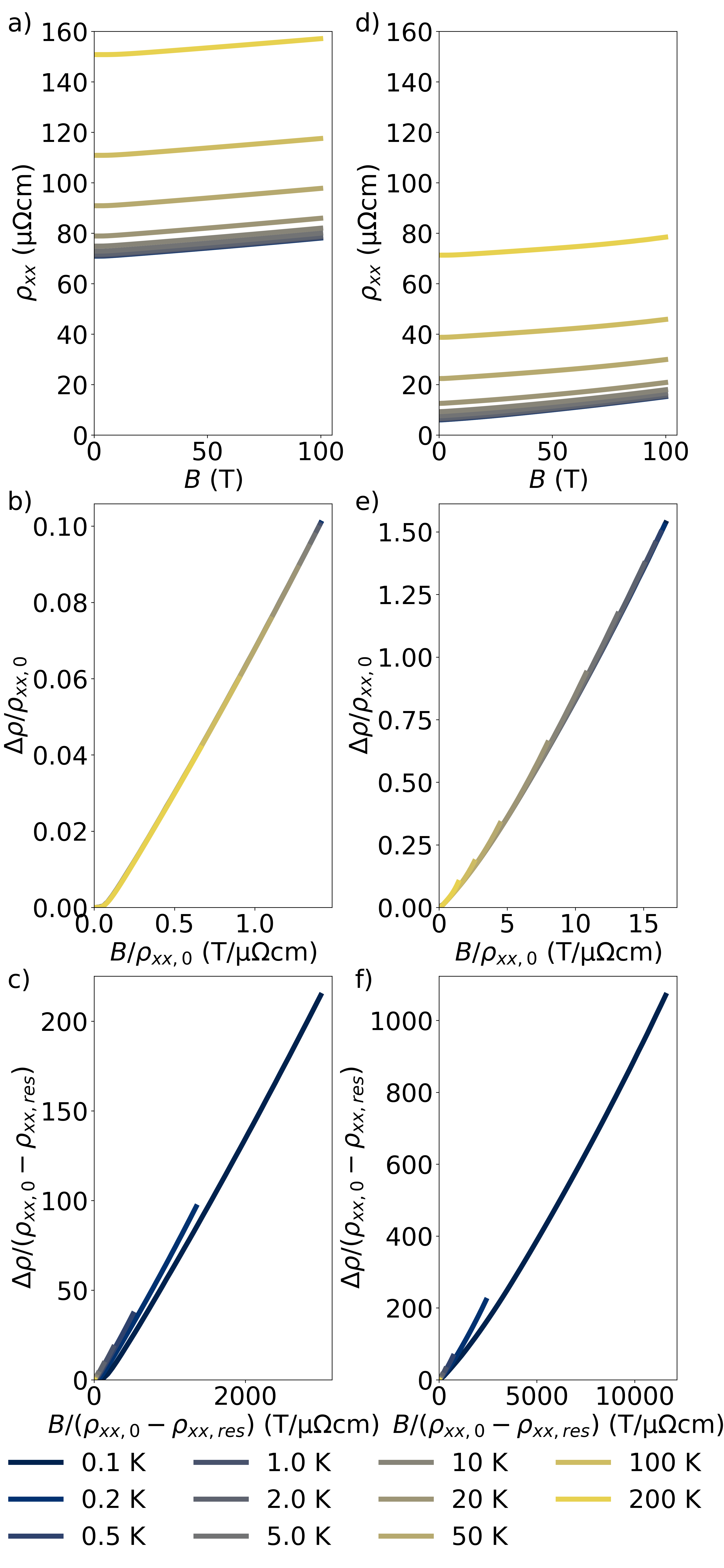}
    \caption{Recovery of Kohler scaling in the presence of a finite residual resistivity. a) Resistivity data obtained in this model, with parameters outlined in the text and a high impurity density of 7$\times10^{16}$ m$^{-2}$. The zero field resistivity satisfies Matthiessen's rule and shows a residual resistivity equivalent to $T_0=177.5$ K extracted using a linear fit to the resistivity. b) Kohler's rule is recovered. c) Kohler's rule without residual resistivity, as in Ref.~\cite{Chan2014} and in quadrature MR, is violated. d)-f) Analogous to a)-c), except with an impurity density of 7$\times 10^{15}$ m$^{-2}$ leading to $T_0=18.4$ K. The high-field deviations in panel e) are discussed in the text. Temperatures are color-coded and shared among all panels.}
    \label{fig:disorder}
\end{figure}

\subsection{Verification of the model}

We start by comparing the model results to Drude theory. In particular, we show in panels a) and b) of Fig.~\ref{fig:disorder_model} the expected exponential distribution of the scattering lifetimes is recovered from the above model, leading to a well defined impurity scattering rate as extracted by the fits. 

Figure \ref{fig:disorder_model}c) furthermore shows the emergence of real-space structure around the impurities. Locally, this results in a breaking of the rotational symmetry $\sigma_{xx}=\sigma_{yy}$, whose patterns are 90 degrees shifted and the symmetry is globally conserved. Intuitively, this results from certain $\phi$ intervals showing a suppressed velocity-velocity correlation in the vicinity of an impurity, breaking the isotropy. Globally, each $\phi$ interval obtains these effects equally through rotation symmetry and $\sigma_{xx}=\sigma_{yy}$ is recovered. Similarly, time reversal symmetry in the absence of a magnetic field guarantees $\sigma_{xy}=0$, which is found to be globally conserved yet locally broken around the (non-magnetic) impurities in Fig.~\ref{fig:disorder_model}d).

We devised further tests to quantitatively recover the Drude result in the absence of boundary scattering, but in the presence of a magnetic field:

\begin{eqnarray}
    \sigma_{xx} &=& \frac{ne^2\tau_0}{m^*} \frac{1}{1+\omega_c^2\tau_0^2} \\
    \sigma_{xy} &=& \frac{ne^2\tau_0}{m^*}  \frac{\omega_c\tau_0}{1+\omega_c^2\tau_0^2} \\
    n &:=& \frac{k_F^2}{2\pi c} \\
    k_F &:=& \frac{m^*v_F}{\hbar} \\
    \omega_c &:=& \frac{qB}{m^*}
\end{eqnarray}

In the above equations, we use $m^*=5m_0$, $v_F=1.68$ m/s and $c=1$ nm, parameters representative of a typical cuprate. These parameters do not influence the conclusions in any way and are kept constant throughout the results shown.

Finally, all conductivity values presented below are the mean of 15 samples, each with a different pseudo-random impurity distribution. These repetitions allow us to obtain a representative mean for $\sigma_{xx}$ as well as to estimate the standard deviation of $\sigma_{xx}$ between samples. The same 15 samples are used as a function of temperature and magnetic field for consistent results. For the results shown in the next subsection, the standard deviation was found to be around 0.5 \%. A small fraction of this variation manifests as a random variation of the macroscopically averaged scattering rate with magnetic field.

\subsection{Results}

After verifying the code, we run the program as a function of temperature and magnetic field. We present here two representative results. First, we find Matthiessen's rule is recovered in zero field as expected. We extract the $T$-linear slope using a fit to the zero-field resistivity and define the effective impurity scattering temperature $T_0$ through $\rho(0)=\alpha(T+T_0)$. 

Using this relation, we find the high impurity case in Fig.~\ref{fig:disorder}a)-c) has $T_0=177.5$ K, similar to what is reported for Bi2201 in Ref.~\cite{Ayres2021}. A snapshots of a single sample at 50 T and 0.1 K where impurity scattering dominates, is shown in panels e3) and e4) of Fig.~\ref{fig:disorder_model}. The region size $L$ is 0.2 $\mu$m and the bulk environment stretches out 1 $\mu$m beyond what is shown to all sides. No charge escaped this bulk region and the mean-free-path is 7.2 nm. Figure \ref{fig:disorder}d)-f) uses instead a more modest and typical impurity scattering equivalent to $T_0=18.4$ K. Snapshots of this scenario are shown in panels e1) and e2) of Fig.~\ref{fig:disorder_model}. $L$ is again 0.2 $\mu$m, but the bulk environment stretches out 10 $\mu$m beyond the real-space patch with a mean-free-path in the zero temperature limit of 69 nm. Again, no charge escaped or made a full cyclotron orbit.

Figures \ref{fig:disorder}b) and \ref{fig:disorder}e) are the primary results, revealing that Kohler scaling is recovered in either case. Fig.~\ref{fig:disorder}e) shows deviations of Kohler scaling at high magnetic fields. We assign this to a numerical instability of the model as $\omega_c\tau_0$ grows. In the high $\omega_c\tau_0$ limit, cyclotron orbits become small and it becomes possible certain orbits do not intersect with any impurity. Even though the presented computations showed no completed cyclotron orbits, a magnetic field dependence of the impurity scattering rate onsets. This could be resolved by using an explicit electric field. Because the underlying scattering rate is no longer given by $\tau_0$, scaling through $\omega_c\tau_0$ fails. Because the magnetic field where these deviations set in scale with $T_0$, no deviations are observed in the high impurity density case.

\begin{figure}[ht!]
    \centering
    \includegraphics[width=0.9\columnwidth]{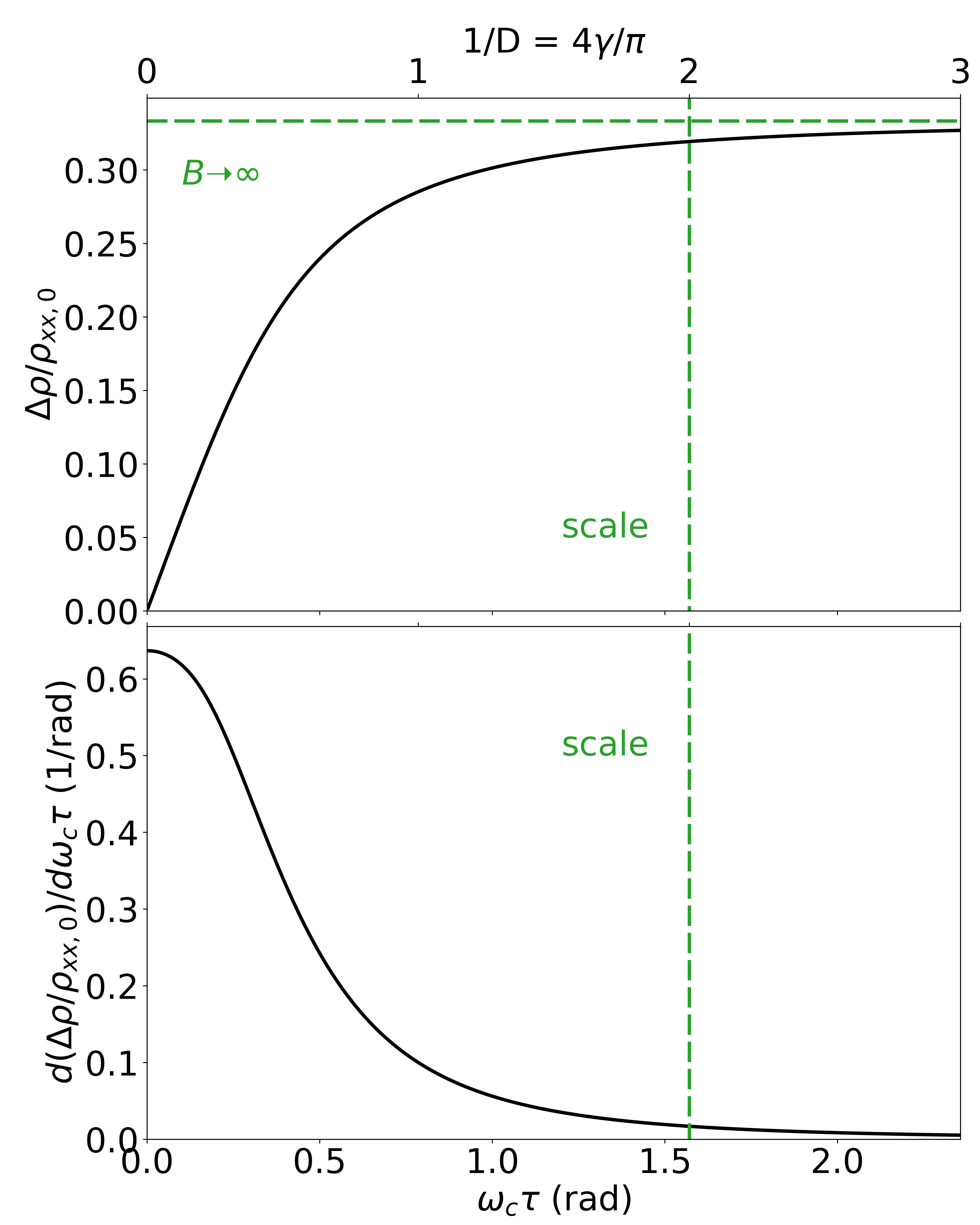}
    \caption{Accidental $B$-linearity in the square Fermi surface model. a) Relative MR given by Eq.~\eqref{squareFS} for a perfectly square Fermi surface. The $\gamma:=\omega_c\tau$ value is normalised such that $2\pi$ corresponds to a full orbit as introduced in Ref.~\cite{Pippard1989}. The low field Taylor expansion is strictly $B$-linear. Horizontal axes are shared between the panels. The green lines correspond to the field scale of the problem in the analytical solution and the saturated MR in the infinite field limit. b) The derivative clearly exposes the accidental nature of the $B$-linear MR at lowest magnetic field.}
    \label{fig:square_fs}
\end{figure}

\section{Accidental $B$-linear MR}
\label{appendix:square_fs}

In this appendix, we highlight an important distinction between impeded cyclotron motion and sharp Fermi surface corners. In the main text, we argued that since the $B$-linear MR obtained from impeded cyclotron motion as well as quadrature MR as defined in Eq.~\eqref{quadrature} are non-accidental. We show here that this is not the case for sharp Fermi surface corners. The square Fermi surface model was originally presented in Ref.~\cite{Pippard1989}. The Fermi surface in question had sides $K$ and a Fermi velocity $v_F$, while the magnetic field $B$ was applied along the symmetry axis. Here, cyclotron motion is given by semi-classical equations of motion and after renormalizing a full orbit to $2\pi$, the effective cyclotron frequency $\gamma$ is defined as:

\begin{eqnarray}
    \gamma &:=& \omega_c\tau = \frac{\pi e v_F}{2K\hbar} B\tau \\
    D &:=& \frac{\pi}{4\gamma}
\end{eqnarray}

 The parameter $2D$ is defined in Ref.~\cite{Pippard1989} as the internal field scale of the problem and corresponds to the magnetic field at which charge can traverse one side of the square (i.e. $\omega_c\tau = \pi/2$). The MR is derived analytically in Ref.~\cite{Pippard1989} and given by:

\begin{eqnarray}
    \frac{\rho_{xx}}{\rho_{xx}(0)} &=& \frac{2D^2 \cosh(2D) - D\sinh(2D)}{(2D^2+1)\cosh(2D)-2D\sinh(2D)-1}
    \label{squareFS} \\
    \lim_{B\rightarrow\infty} && \frac{\rho_{xx} - \rho_{xx}(0)}{\rho_{xx}(0)} = \frac{1}{3} \\
    \lim_{B\rightarrow0} && \frac{\rho_{xx} - \rho_{xx}(0)}{\rho_{xx}(0)} = \frac{2}{D} = \frac{2}{\pi}\omega_c\tau 
\end{eqnarray}

This relation is shown in Figure \ref{fig:square_fs}. From the derivative to effective magnetic field, it is apparently clear that the $B$-linear MR obtained in this way is accidental. Even at field scales much below $\pi/2$, deviations from linearity are visible. Softening of the corners introduces a $B^2$ MR at low field until $\omega_c\tau$ is sufficient such that carriers can traverse the corner, but cannot resolve the accidental nature of the $B$-linear MR. 

Recently, sharp Fermi surface corners \cite{Feng2019} and turning points \cite{Koshelev2013, Maksimovic2020} have been proposed to explain the $B$-linear MR in ordered phases of either charge or spin density waves. We urge caution in studying such Fermi surface features in isolation and recommend studying the onset of saturation through inter-corner effects. The saturation field scale is important to estimate when quantum oscillations can be expected, as well as to make an independent measurement of $\tau/m^*$ on the flat Fermi surface sections. 

The importance of this saturation to experiment can be readily demonstrated. The saturation in Fig.~\ref{fig:square_fs} amounts to 33 \% MR, whereas the parent compound Ba122 is known to show unsaturated $B$-linear MR to above 20 \% \cite{Maksimovic2020, Huynh2011}. For CDW materials, this limit is known to be widely violated \cite{Feng2019}. This indicates saturation would have been expected to set in. One consideration is that the turning points or corners are subject to a high density of states and could manifest in locally enhanced scattering rates or masses, which would lower the magnetic field at which the saturation can be observed. Of course, these considerations neglect the multi-band character of Ba122 or various density wave materials, but justifies further investigation into shorting effects and multi-band character in such systems.

\end{document}